\documentclass[review]{elsarticle}

\usepackage[top=50mm, bottom=50mm, left=50mm, right=50mm]{geometry}
\usepackage{lineno}
\usepackage{amssymb}
\usepackage{amsmath}
\usepackage{amsthm}
\usepackage{epsfig}
\usepackage{graphicx}
\usepackage{graphics}
\usepackage{float}
\usepackage{subcaption}
\usepackage{multirow}
\usepackage{color}
\usepackage{lineno}
\usepackage{fullpage}
\usepackage[normalem]{ulem} 
\usepackage{makeidx}
\usepackage{xspace}
\usepackage{wrapfig}
\usepackage{xcolor}
\usepackage{url}
\usepackage{booktabs}
\usepackage{todonotes}

\definecolor{darkgreen}{rgb}{0.0, 0.2, 0.13}
\definecolor{burgundy}{rgb}{0.5, 0.0, 0.13}

\newcommand\REV[1]{{\color{black} #1}}
\newcommand\RED[1]{{\color{black} #1}}

\newcommand\LD[1]{{\color{black} #1}}
\newcommand\NP[1]{{\color{black} #1}}

\modulolinenumbers[5]

\everymath{\displaystyle}

\begin{document}

\begin{frontmatter}

\title{Modelling the COVID-19 \LD{epidemic} and the vaccination campaign in Italy by the SUIHTER model}
\author[1]{Nicola Parolini}
\author[1]{Luca Dede'}
\author[1]{Giovanni Ardenghi} 
\author[1,2]{Alfio Quarteroni}
\address[1]{MOX, Department of Mathematics, Politecnico di Milano, Italy}
\address[2]{Institute of Mathematics, \'Ecole Polytechnique F\'ed\'erale de Lausanne
(EPFL), Switzerland (Professor Emeritus)}

\begin{abstract}
Several epidemiological models have been proposed to study the evolution of COVID-19 pandemic. In this paper, we propose an extension of the \texttt{SUIHTER} model, first introduced in [Parolini et al, Proc R. Soc. A., 2021]  to analyse the COVID-19 spreading in Italy, \LD{which} accounts for the vaccination campaign and the presence of new variants when they become dominant. In particular, the specific features of the variants (e.g. their increased transmission rate) and vaccines (e.g. their efficacy to prevent transmission, hospitalization and death) are modeled, based on clinical evidence. The new model is validated comparing its near-future forecast capabilities with other epidemiological models and exploring different scenario analyses.
\end{abstract}


\end{frontmatter}

\section{Introduction}

Since the beginning of the COVID-19 pandemic in early 2020, the mathematical community has devoted an unprecedented effort in developing {novel} mathematical, numerical and statistical tools to investigate the evolution of the most challenging public health threat {in decades}. For an overview on recent advances 
\REV{in mathematical epidemiology, computational modeling, physics-based simulation, data science, and machine learning applied to the COVID-19 pandemic}, {we refer the reader to} \cite{kuhl2021computational}.
Even limiting the focus to the Italian \LD{context,} several contributions have been proposed to accurately describe the {spatio-temporal spreading} of the epidemic in Italy \cite{sidarthe,Gatto,Bertuzzo,DellaRossa,10.1371/journal.pone.0237417}, to forecast its future evolution \cite{farcomeni2020,parolini2021suihter,mira2021} and to quantify (and possibly optimize) the effects of containing measures, including both pharmaceutical and non-pharmaceutical interventions (NPIs) \cite{Marzianoe2019617118,bonifazi2021study,Giordano2021}.

\LD{As for every other virus, SARS-CoV-2, the virus responsible of COVID-19, changes in time through mutations that generate variants.} Variants may differ from the wild-type virus by transmission rates, capability of leading to severe disease, response to vaccination. Given the enormous impact that the COVID-19 pandemic had (and is still having) on global health, economy, environment and society, the appearance and evolution of SARS-CoV-2 have been carefully tracked since the beginning of the pandemic by collecting sequences on shared public repositories \cite{pango,nextstrain,gisaid} and monitoring, in particular, the so-called \textit{variants of interest} (VOI) and \textit{variants of concern} (VOC). In particular, the latter are defined as those variants which may represent an additional risk due to an increased {transmissibility}, an increased virulence or a decrease in the effectiveness of public health and social measures or available diagnostics, vaccines, therapeutics. Since the beginning of the pandemic, four variants have been classified as VOC by the World Health Organization \cite{variantWHO}, namely the \textit{Alpha} variant (lineage B.1.1.7), designated on December 18, 2020, first documented in United Kingdom in September 2020, the \textit{Beta} variant (lineage B.1.351), designated on December 18, 2020, first documented in South Africa in May 2020, the \textit{Gamma} variant (lineage P.1), designated on January 11, 2021, first documented in Brazil in November 2020, and  the \textit{Delta} variant (lineage B.1.617.2),  designated on May 11, 2021, first documented in India in October 2020. \RED{Finally, the recent variant \textit{Omicron} was first documented in November 24, 2021 in South Africa and designated on November 26, 2021. It is currently subject of an intense effort by the scientific community to characterize its behaviour in terms of transmissibility, disease severity, immune escape.} 

When the \REV{prevalence} of one (or more) of these variants become relevant, the epidemiological models used to track the pandemic evolution should take them into account. Several modeling contributions to describe the effect of variants have been proposed in the literature in the past few months \cite{Giordano2021,RAMOS2021105937,Krueger2021.05.07.21256847,APandemicForecastingFrameworkAnApplicationofRiskAnalysis}. 

Another relevant contribution that deserves the attention of the modelling community is the impact of the vaccination campaign that has now reached, in \LD{several countries,} a large coverage of the population. Different vaccines for COVID-19 have been authorized by the public health organizations and introduced in the market since late 2020. In particular, in the western countries, the most of the vaccination campaign have been carried out using four products: the \textit{Comirnaty} vaccine (from Pfizer/BioNTech), the \textit{Spikevax} vaccine (from Moderna), the \textit{Vaxzevria} vaccine (from AstraZeneca) vaccine, and the \textit{Janssen} vaccine (from Johnson \& Johnson). The first two (Comirnaty and Spikevax) are based on the messenger RNA (mRNA) technology and \LD{similarly behave} in terms of efficacy \cite{SelfEtAl2021}. AstraZeneca and Janssen vaccines are based on the more traditional viral vector technology. Among the four vaccines, Janssen is the only proposed (at least in the first phase of the vaccination campaign) with a single dose administration, while the others are supplied with two doses separated by a time delay ranging between 3 and 6 weeks for mNRA vaccines and between 4 and 12 weeks for Vaxzevria.

SUIHTER is an epidemiological model, first introduced in \cite{parolini2021suihter}, {which is designed to conform to} the epidemiological data that have been made available by the Italian Authorities since the beginning of the epidemics. The choice of the compartments defining the model has been driven by the assumption that the model should match as close as possible \LD{such} available data. 

In this paper, we propose an extension of the SUIHTER model able to include both the effect of the vaccination campaign and of the two variants that appeared and became dominant in Italy during \LD{2021}, the \textit{Alpha} variant (between February and March 2021) and the \textit{Delta} variant (between July and August 2021). {Moreover, coherently with the principle that the model should conform as much as possible to available data, it has also been adapted to match additional information that meanwhile became available from authorities, specifically some additional fluxes between compartments (such as new admission in hospital and ICUs).}  \LD{The analysis does not consider the \textit{Omicron} variant due to the currently limited knowledge about its increased transmissibility and low prevalence at present time (mid-December 2021).}

Several attempts have been made to propose rigorous evaluation strategies to compare the forecast capabilities of different models; \LD{see, e.g., \cite{Bracher2021,Cramer2021.02.03.21250974}.} Ensemble approaches combining multiple available forecasts from different models have proved to outperform single models in forecasting the early phases of COVID-19 pandemic \cite{Bracher2020.12.24.20248826,Funk2020.11.11.20220962,TAYLOR2021}, as well as influenza \cite{Reich3146,Reich2019}, dengue fever \cite{Johansson2019}, and Ebola \cite{VIBOUD201813} outbreaks. The superior performance of ensemble forecast have been shown in different disciplines and they are typically associated to the ability of integrating the information coming from different models producing accurate predictions with well-calibrated uncertainty \cite{doi:10.1057/jors.1969.103,LEUTBECHER20083515}, avoiding, at the same time, the risk of delayed (or missing) forecasts or premature interruption of the forecast supply, which may occur when relying on a single specific model. 

To evaluate its performance in terms of forecasting capability, the SUIHTER model has joined the European COVID-19 Forecast Hub \cite{forecasthub} since April 2021. The aim of this collaborative project is to provide decision makers and the general public with reliable estimates of the near-term future evolution of the COVID-19 pandemic in European countries by collecting forecasts from different models into an ensemble. \REV{A quantitative comparison based on suitable ranking indicators has been carried out, showing that the SUIHTER model can produce reliable short term forecast on different epidemic quantities of interest.}

Moreover, different analyses have been carried out to evaluate at which extent the extended SUIHTER model accounting for variants and vaccination is able to better identify specific epidemic trends (such as new outbreak associated to the  growth of a more transmissible virus variant) and how it can be used to quantify the effect of the vaccination campaign (including the effect of the vaccination rate).

The paper is organized as follows: in Section \ref{sec:model} the SUIHTER model including the vaccine compartment is introduced, and it is later extended in Section \ref{sec:variant_model} to also account for an emerging variant. The strategy used to model multiple scenarios characterized by different NPIs is presented in Section \ref{sec:scenarios}. The numerical results are presented and discussed in Section \ref{sec:results}, including a quantitative validation of the forecasting capabilities of the model and different analyses highlighting the role of the different new features of the model (including new variants and vaccination effects). Finally, in Section \ref{sec:conclusions}, we draw our conclusions and we discuss model’s limitations and some possible future developments.

\section{Mathematical model}\label{sec:model}
The design of the original \texttt{SUIHTER} model, introduced in \cite{parolini2021suihter}, relied on the criterion that the epidemiological compartments should match as close as possible the data that were daily made available by the Italian authorities. In the present work, we propose an extended version of the SUIHTER model which is able to exploit a set of  additional data that started to be released at a later stage (including, e.g., the number of new admissions in hospitals and ICUs, as well as data classified by age and clinical status). Moreover, the new model accounts for the role of the vaccination campaign that is ongoing in most countries since the end of 2020, by adding \NP{three additional compartments collecting vaccinated individuals}.

\begin{figure}[t]
\centering
\includegraphics[width=0.8\textwidth]{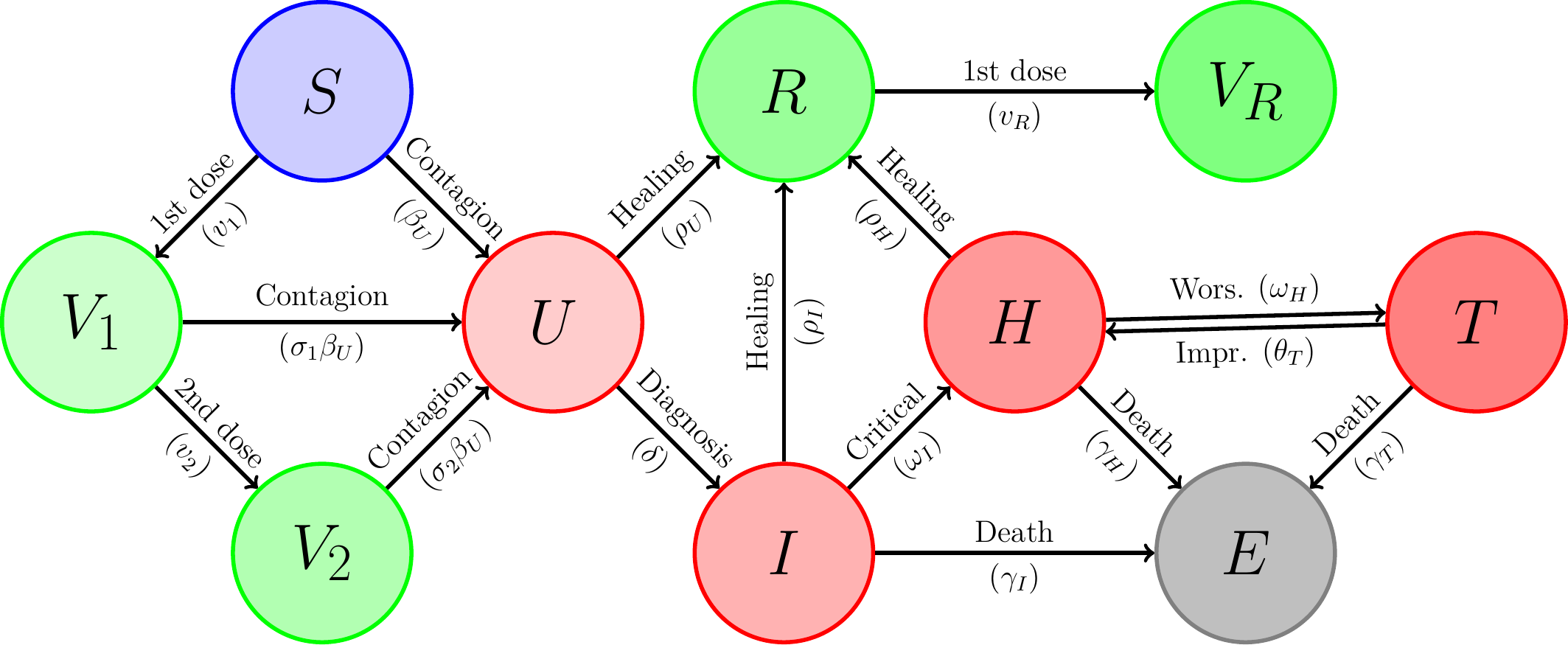}
\caption{Interactions among compartments in \LD{the extended} {\tt SUIHTER} model }
\label{fig:suihtervv}
\end{figure}

The extended \texttt{SUIHTER} model is defined by the following system of ordinary differential equations:
\begin{equation} \label{eq:suihtervv}
    \begin{array}{l}
\dot S(t) = - S(t)\, \frac{\beta_U \, U(t)}{N} - v_1, \\[3mm]
\dot U(t) =   \left( S(t) + \sigma_1 \, V_1(t) + \sigma_2 \, V_2(t) \right) \, \frac{\beta_U \, U(t)}{N} - (\delta + \rho_U)\, U(t), \\[3mm]
\dot I(t) = \delta \, U(t) - (\rho_I + \omega_I +\gamma_I )\, I(t),\\[3mm]
\dot H(t) = \omega_I \, I(t) - (\rho_H + \omega_H + \gamma_H) \, H(t) + \theta_T \, T(t), \\[3mm]
\dot T(t) = \omega_H \, H (t)  - (\theta_T + \gamma_T)\, T(t), \\[3mm]
\dot E(t) = \gamma_I \,  I (t) +\gamma_H \, H(t)+\gamma_T \, T(t), \\[3mm]
\dot R(t) =  \rho_U \, U (t) + \rho_I \, I(t) + \rho_H \, H(t) - \NP{v_R}, \\[3mm]
\dot V_1(t) = v_1 - v_2 -\sigma_1 \, V_1(t) \, \frac{\beta_U \, U(t)}{N}, \\[3mm]
\dot V_2(t) = v_2 -\sigma_2 \, V_2(t) \, \frac{\beta_U \, U(t)}{N}, \\[3mm]
\NP{\dot V_R(t) = v_R,} \\[3mm]
\end{array}
\end{equation}
endowed with suitable initial conditions, where the model compartments are defined as {follows} (see Figure~\ref{fig:suihtervv}):
\begin{itemize}
\item $S$: number of \textit{susceptible} (uninfected) individuals;
\item $U$: number of \textit{undetected} (both asymptomatic and symptomatic) infected individuals;
\item $I$: number of infected individuals \textit{isolated} at home;
\item $H$: number of infected \textit{hospitalized} individuals;
\item $T$: number of infected \textit{threatened} individuals hosted in ICUs;
\item $E$: number of \textit{extinct} individuals;
\item $R$: number of \textit{recovered} individuals;
\item $V_1$: number of individuals partially \textit{vaccinated} with only one dose;
\item $V_2$: number of individuals fully \textit{vaccinated} with two doses;
\NP{\item $V_R$: number of \textit{recovered} individuals who have been \textit{vaccinated},}
\end{itemize}
\noindent and $N=S+U+I+H+T+E+R+V_1+V_2+V_R$ denotes the total population (assumed constant). Here, $t$ measures time in days, compartments report figures in units. \NP{The system is numerically solved with the explicit fourth-order Runge-Kutta scheme with time step $\Delta t=1$ day.}

The quantity $v_1$ denotes the number of susceptible individuals per day who acquire immunity from the first dose. Similarly, $v_2$ denotes the number of individuals per day which are already vaccinated with one dose and acquire a stronger immunity by the second dose. \NP{Moreover, $v_R$ denotes the number of recovered individuals per day who receive a vaccine dose.} The number of first doses $w_1(t)$ and second doses $w_2(t)$ administrated in Italy are daily recorded on the Italian Government repository \cite{vaccdata}. \NP{The quantity $w_1$ does not include the number of vaccine doses $w_R(t)$  administrated to (detected) recovered individuals.}
Unfortunately, these vaccination rates cannot be directly used in system \eqref{eq:suihtervv} since during the whole epidemic period a great number of cases were not identified so that many recovered individuals have been vaccinated as if they were still susceptible.
The number of individuals who recovered before (or without) being detected can be defined as $R_U=R-R_D$, where
$$R_D(t)=\int_{t_0}^t \left(\rho_I I(\tau)+\rho_H H(\tau)\right)\,d\tau,$$
is the number of individuals who recovered after having been detected and can be obtained by postprocessing the computed \LD{compartments $I$ and $H$. 
The indicator $R_U$} can be used to estimate the rate of vaccine doses that have been actually administrated to susceptible individuals, by assuming that recovered individuals who were not formerly detected have the same probability of being vaccinated as susceptible individuals. Namely, the rate of first doses administrated to susceptible individuals are estimated as 
$$\LD{v_1(t) = w_1(t-t_{im}) \, \frac{S}{S+R_U},}$$
where we have assumed that the time required to develop the immunity \LD{is $t_{im}=$ 14 days.}
Similarly, the second doses administrated to partially vaccinated individuals ($V_1$) are only a fraction of the total number of second doses, therefore we get 
$$\LD{v_2(t) = w_2(t-t_{im}) \, \frac{S}{S+R_U}.}$$ 
\NP{Finally, the total number of doses per day that recovered individuals (either detected or undetected) receive is given by:
$$v_R(t) = w_R(t-t_{im}) + w_1(t-t_{im}) \, \frac{R_U}{S+R_U},$$}


The model can be initialized at any time $t_0$ prior to the start of the vaccination campaign (December 20, 2020 in Italy) by using the data (for those compartments for which data are available, namely $I,H,T,E$), null initial values for $V_1$ and $V_2$, while, as proposed in \cite{parolini2021mathematical}, the \textit{Undetected} and \textit{Recovered} compartments are initialized as
$$
R(t_0) = \left( \frac{1}{\text{IFR}(t_0)} - 1\right)\, E(t_0), \quad \quad
U(t_0) = \left( \frac{\text{CFR}(t_0+d)}{\text{IFR}(t_0)} - 1 \right)\, \left( I(t_0)+H(t_0)+T(t_0) \right),
$$
where IFR$(t)$ is the \textit{Infection Fatality Ratio}, CFR$(t)$ is the time-dependent \textit{Case Fatality Ratio} and $d=13$ days denotes the confirmation-to-death delay.
\RED{In \cite{parolini2021mathematical}, we considered a constant IFR computed as the weighted average of the age-specific IFR estimates weighted \LD{by the population age structure, under the assumption of equal attack rates across age-groups, as proposed in \cite{imperialreport}.}
A better estimate of the reference IFR$(t)$ at a specific time can be computed by considering the variable percentage of each age-group among the total infected over time (data available on \cite{ISSdata} since December 8, 2020), namely:
$$\text{IFR}(t)=\sum_{i=1}^m q_i(t) \, \text{IFR}_i ,$$
where IFR$_i$ denotes the infection fatality ratio for age-group $i$ and $q_i(t)$ is the percentage of infected at time $t$ belonging to age-group $i$. This correction may be particularly relevant when using the model in a time frame in which the vaccination campaign is not homogeneous across age-groups, so that the percentage of different age-groups among the total infected may vary significantly. A further improved estimate for IFR$(t)$ can be obtained considering the vaccine effectiveness in reducing the mortality for different age-groups, namely:
\begin{equation}\label{eq:IFR}
\text{IFR}(t)=\sum_{i=1}^M q_i(t)\,  [(1-v_{i,1}(t)-v_{i,2}(t))+v_{i,1}(t)\, m_{i,1}+v_{i,2}(t) \, m_{i,2}] \, \text{IFR}_i,
\end{equation}
where $v_{i,1}(t)$ and $v_{i,2}(t)$ denote the \LD{fraction of individuals in} age-group $i$ who received one or two doses, respectively, at time $t$, while $m_{i,1}$ and $m_{i,2}$ denote the reduced probability to die for an isolated infected individual of age-group $i$ who is vaccinated (with one or two doses, respectively)  w.r.t. an unvaccinated isolated infected individual in the same age-group. 

The time-history of the CFR$(t)$ since the beginning of the epidemic is shown in Figure~\ref{fig:IFR} (left) where the high values during the first epidemic wave clearly indicates the strong underreporting. In Figure~\ref{fig:IFR} (right),  CFR$(t)$ and reference IFR$(t)$ estimated with \eqref{eq:IFR} (which can be computed based on data available from December 8, 2020) are compared. The effect of the vaccination campaign during year 2021 is clearly visible in the reduction of the reference IFR$(t)$.   

\begin{figure}
    \centering
    \subcaptionbox{CFR}{\includegraphics[width=0.49\textwidth, trim={80 60 80 60}]{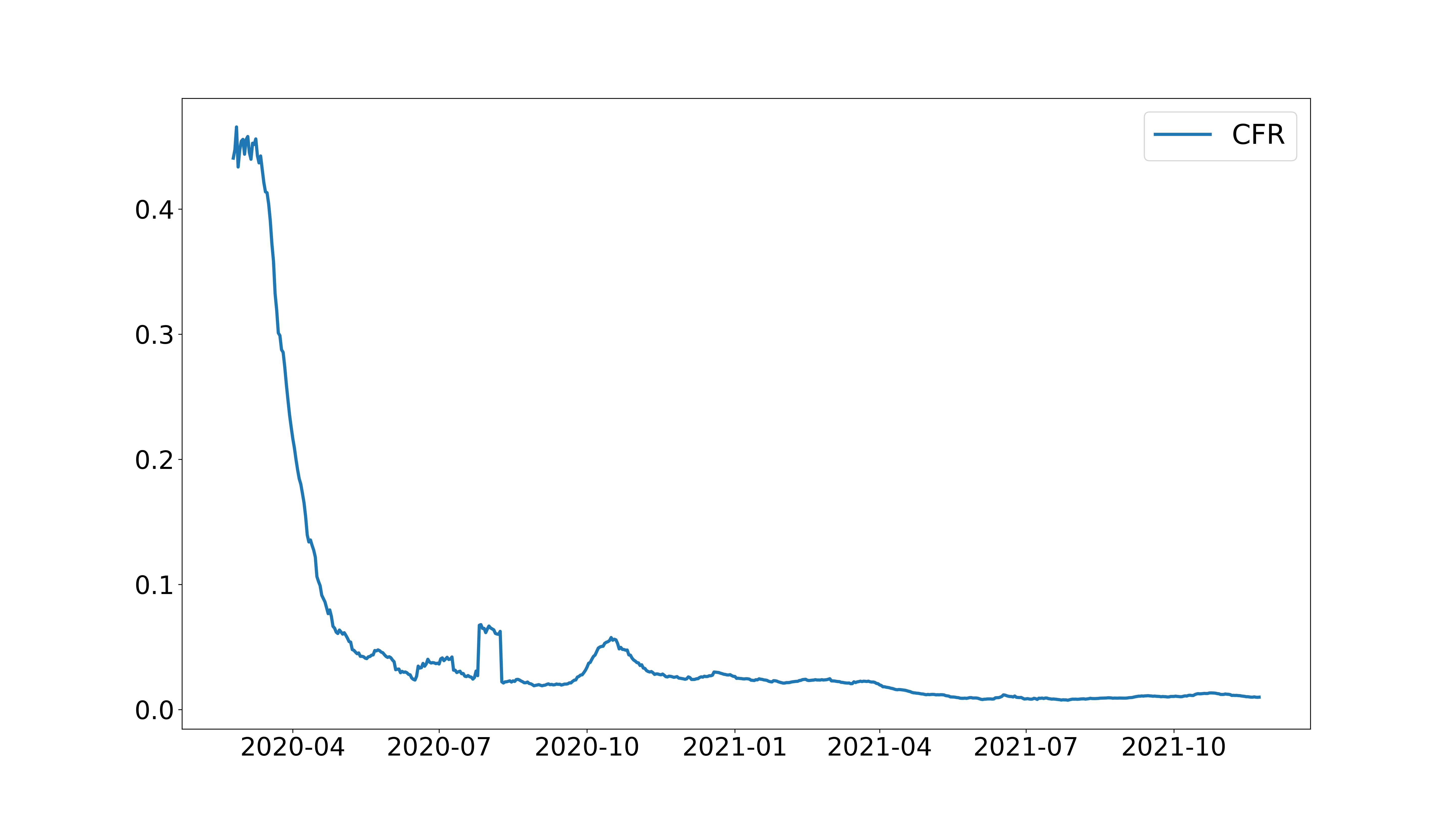}}
    \subcaptionbox{CFR vs IFR}{\includegraphics[width=0.49\textwidth, trim={70 60 80 60}]{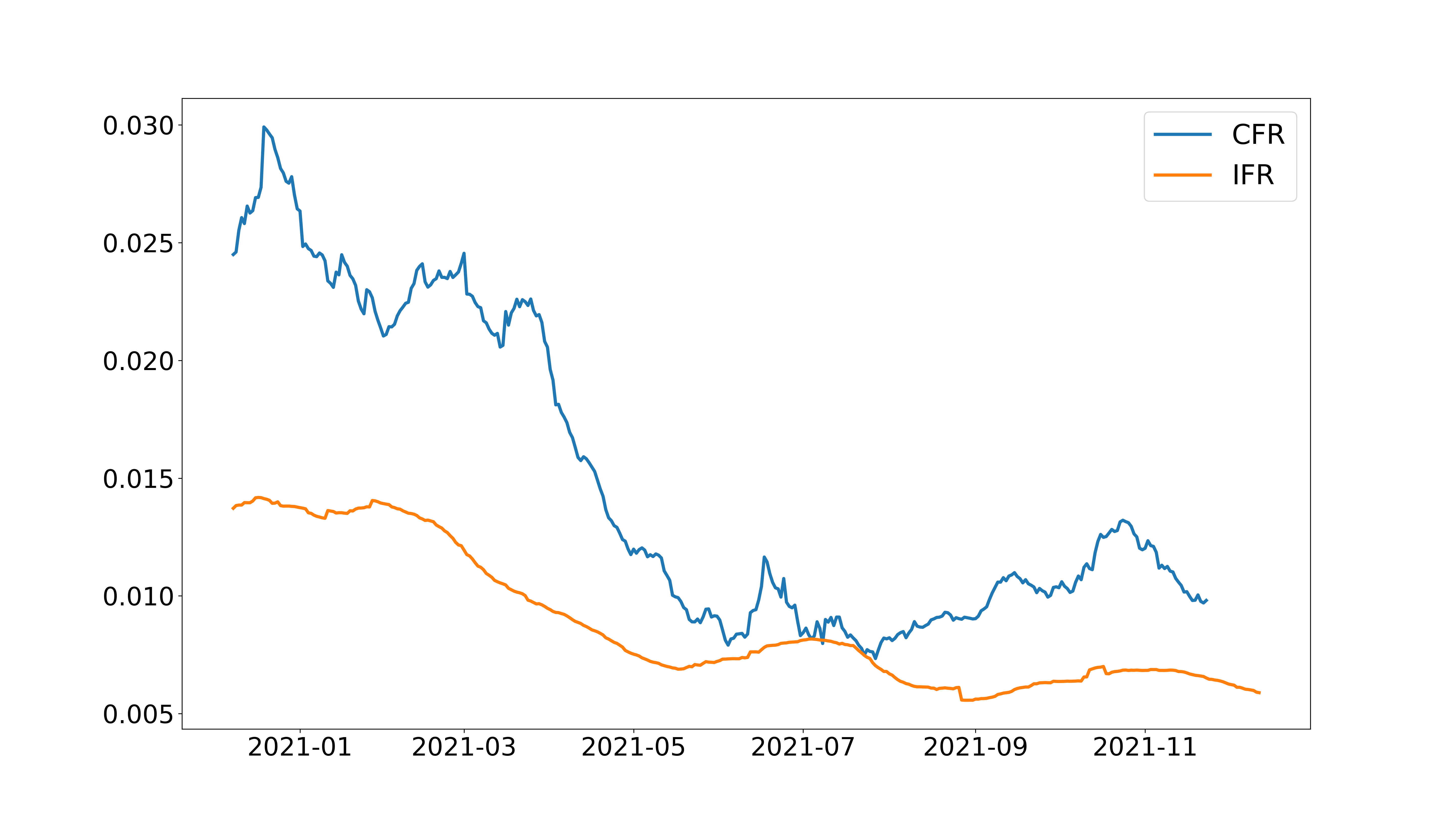}}
    \caption{Time-evolution of the reference CFR$(t)$ since the beginning of the epidemic (left) and a comparison between CFR$(t)$ and IFR$(t)$ accounting for age-dependent attack rates and vaccine efficacy in preventing death since December 2020 (right)}
    \label{fig:IFR}
\end{figure}
}

Model \eqref{eq:suihtervv} depends on the following \LD{parameters (rates)}:
\begin{itemize}
\item $\beta_U$ denotes the transmission rate due to contacts between susceptible and undetected infected individuals;
\item $\omega_I$ denotes the rate at which $I$-individuals develop clinically relevant symptoms, while $\omega_H$
denotes the rate at which $H$-individuals develop life-threatening symptoms;
\item $\theta_T$ denotes the rate at which $T$-individuals improve their health conditions and return to the less critical $H$ compartment;
\item $\delta$ denotes the rate of detection, relative to undetected infected individuals;
\item $\rho_U$, $\rho_I$ and $\rho_H$ denote the rate of recovery for three classes ($U$, $I$ and $H$, respectively) of infected subjects;
\item $\gamma_I$, $\gamma_H$ and $\gamma_T$ denote the mortality rates for the individuals isolated at home, hospitalized and hosted in ICUs, respectively.
\end{itemize}

Some of the parameters in system \eqref{eq:suihtervv} are evaluated starting from the available data. In particular, the worsening rates $\omega_I$ and $\omega_H$ are obtained directly from the new hospitalization and the new admission data as follows
\begin{equation}\label{eq:worsening}
    \omega_I(t) = \frac{\hat{H}^+(t)}{\hat{I}(t)}, \qquad 
    \omega_H(t) = \frac{\hat{T}^+(t)}{\hat{H}(t)}, 
\end{equation}
\noindent where $\hat{H}$ are the daily data provided by the Dipartimento della Protezione Civile (DPC) for hospitalized individuals, while $\hat{H}^+$ and $\hat{T}^+$ are the daily data provided by Istituto Superiore di Sanità (ISS) \cite{ISSdata} and DPC \cite{PCM-DPC} respectively for new hospitalizations and admissions in ICUs.
To reduce the effects of the weekly oscillation associated to irregular data reporting, the values of  $\omega_I$ and $\omega_H$ used in \eqref{eq:suihtervv} are the 1-week moving averages of the time series estimated in (\ref{eq:worsening}). 

The detection rate $\delta$ is also estimated by assuming that at each time the probability $p_D$ of being detected can be related to the ratio between the infection fatality rate and the time-dependent case fatality rate, namely
$${p_D(t)=\frac{IFR(t)}{CFR(t+d)}.}$$
Assuming a mean detection time $t_D=8$ days, the detection rate is then computed at each time $t$ as
$$\delta(t) = \frac{p_D(t)}{t_D}.$$
To avoid spurious oscillations, the time-history of $p_D(t)$ is also smoothed taking its weekly moving average.
\RED{The time-history of the probability $p_D$ of being detected at time $t$, based on the estimate of IFR$(t)$  given in \eqref{eq:IFR}, is displayed in Figure \ref{fig:delta} where we can notice that the at mid July 2021, when the lowest incidence was achieved, the probability of being detected was almost 100\%. 

\begin{figure}
    \centering
    \includegraphics[width=\textwidth, trim={100 80 100 80}]{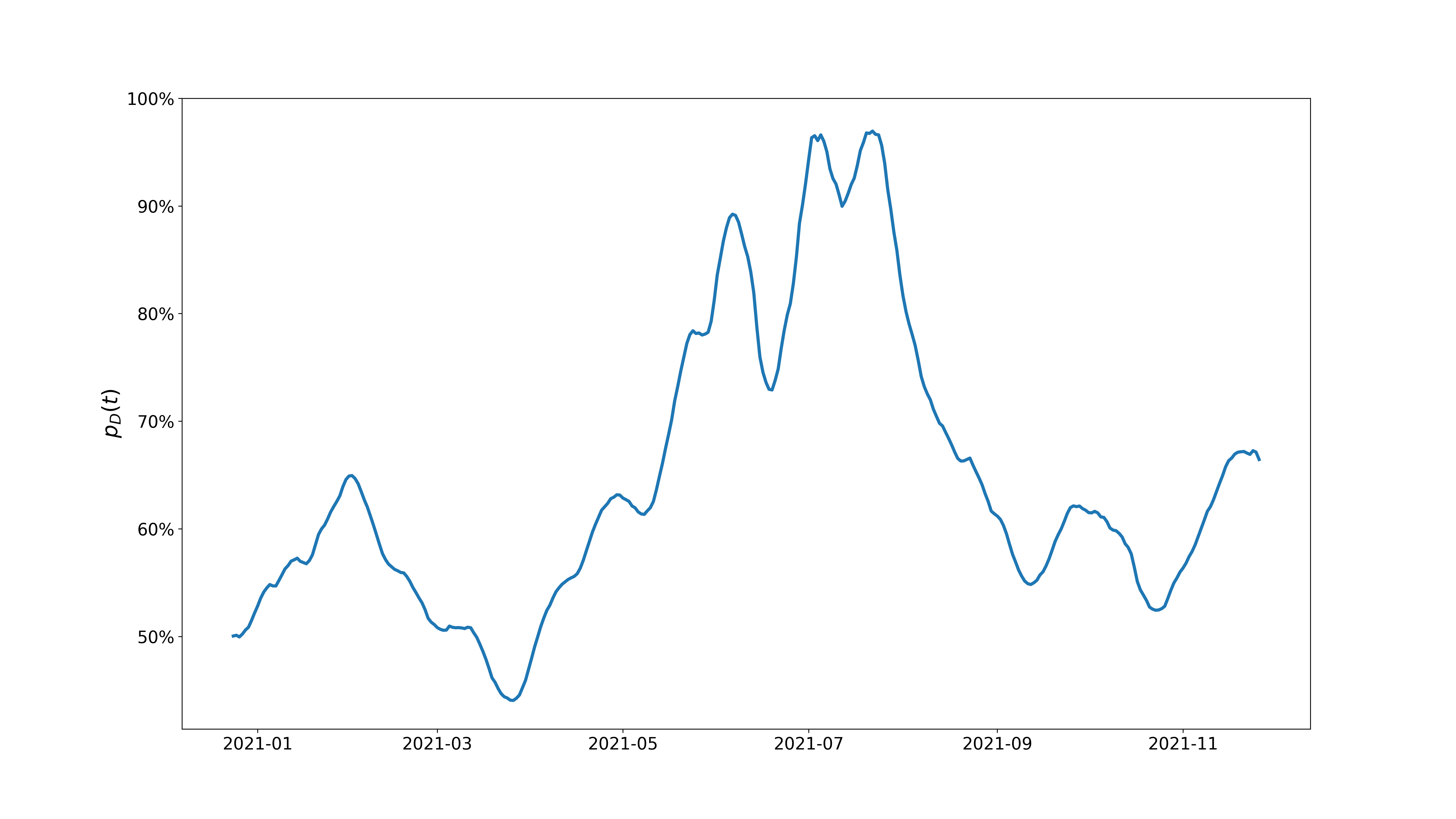}
    \caption{Time-evolution of the probability $p_D$ of being detected}
    \label{fig:delta}
\end{figure}
}

Given an estimate of $\delta$, if we assume that all \textit{Undetected} individuals can leave the compartment only by either being detected or recovering, we can also estimate the \textit{Undetected} recovery rate $\rho_U$ as 
$$\rho_U(t) = \frac{1 - t_D \, \delta(t)}{t_R},$$
where $t_R$ is the mean recovery time for $Undetected$, which is estimated through the calibration process.

At this stage, the effect of the vaccine in the model \RED{is given by the susceptibility of vaccinated individuals with respect to non-vaccinated individuals, given by the parameters $\sigma_1=0.3$ and $\sigma_2=0.12$ \cite{monitoraggioISS} for partially and completely vaccinated individuals, respectively.}

\LD{The remaining} parameters of the model need to be calibrated  to reproduce the epidemic history. Some parameters \LD{can be considered as} constant in time throughout the pandemic, \LD{while we prescribe the other ones} are piecewise constant over $n$ time phases $[t_k,t_{k+1}]$, $k=1,\dots,n$ of prescribed duration (typically equal to 15 days) in which the simulation time windows $[t_0,t_F]$ is subdivided.
The parameters chosen to be constant in time are: $t_R$, $\gamma_I$ and $\gamma_T$, while $\beta_{U,k}$, $\rho_{I,k}$, $\rho_{H,k}$, $\theta_{T,k}$ and $\gamma_{H,k}$, $k=1,\dots,n$ are constant on the $k$-th phase $[t_k,t_{k+1}]$.
All these parameters are calibrated through a two-step calibration: a first estimate is obtained with a least square procedure, this estimate is then used as initial condition for a Monte Carlo Markov Chain (MCMC) procedure. The prior is a uniform distribution around the least square estimate with $\pm30\%$ of the parameter value interval. For a more detailed description of the calibration process, see \cite{parolini2021suihter}.

\subsection{Accounting for virus variants}\label{sec:variant_model}
With the appearance of virus variants featuring a higher transmission rate than the wild type virus \cite{campbell2021increased},  predicting the evolution of the epidemic curve has become more challenging. As a matter of fact, when a more aggressive variant starts spreading, the epidemiological characteristic could change, even drastically. When using the model for forecasting the future evolution of the epidemic in the presence of an emerging variant, it is necessary to include in model \eqref{eq:suihtervv} the effects of variants. In particular, the model is modified by splitting the \textit{Undetected} compartment, in which the contagion occurs, into two new compartments:
\begin{itemize}
    \item $U^b$: number of \textit{Undetected} (both asymptomatic and symptomatic) infected with the base variant, that is the variant being dominant at the initial time of the simulation,
    \item $U^v$: number of \textit{Undetected} (both asymptomatic and symptomatic) infected with the variant of concern that is spreading faster among the population. 
\end{itemize}

Susceptible individuals have different probabilities of being infected by people belonging to each of these two compartments, typically the new variant has a higher transmission rate than the base one. For example the \textit{Delta} variant has been estimated to be around $50\%$ more transmissible than the \LD{\textit{Alpha}} one \cite{deltaVariant}, and, in turn, it was previously estimated to be $38\%$ more transmissible than the wild type virus \cite{tanaka2021increased}.

\LD{The model accounting for a virus variant reads:}
\begin{equation}\label{eq:suihtervv_forecast}
    \begin{array}{l}
\dot S(t) = - S(t)\, \frac{\beta_U^b \, U^b(t) +\beta_U^v \, U^v(t)}{N} - v_1, \\[3mm]
\dot U^b(t) =   \left( S(t) + \sigma_1^b \, V_1(t) + \sigma_2^b \, V_2(t) \right) \, \frac{\beta_U^b \, U^b(t)}{N} - (\delta + \rho_U)\, U^b(t), \\[3mm]
\dot U^v(t) =   \left( S(t) + \sigma_1^v \, V_1(t) + \sigma_2^v \, V_2(t) \right) \, \frac{\beta_U^v \, U^v(t)}{N} - (\delta + \rho_U)\, U^v(t), \\[3mm]
\dot I(t) = \delta \, (U^b(t) + U^v(t)) - (\rho_I + \tilde{\omega}_I + \Tilde{\gamma}_I)\, I(t),\\[3mm]
\dot H(t) = \Tilde{\omega}_I\, I(t) - (\rho_H + \Tilde{\omega}_H + \gamma_H) \, H(t) + \theta_T \, T(t), \\[3mm]
\dot T(t) = \Tilde{\omega}_H \, H (t)  - (\theta_T + \gamma_T)\, T(t), \\[3mm]
\dot E(t) = \Tilde{\gamma}_I \, I (t) +\gamma_H \, H(t)+\gamma_T \, T(t), \\[3mm]
\dot R(t) =  \rho_U \, (U^b (t)+U^v(t)) + \rho_I \, I(t) + \rho_H \, H(t) - \NP{v_R}, \\[3mm]
\dot V_1(t) = v_1  - v_2  -\sigma_1^b \, V_1 \, \frac{\beta_U^b \, U^b(t)}{N} -\sigma_1^v \, V_1 \, \frac{\beta_U^v \, U^v(t)}{N}, \\[3mm]
\dot V_2(t) = v_2 -\sigma_2^b \, V_2 \, \frac{\beta_U^b \, U^b(t)}{N} -\sigma_2^v \, V_2 \, \frac{\beta_U^v \, U^v(t)}{N}, \\[3mm]
\NP{\dot V_R(t) = v_R,} \\[3mm]
\end{array}
\end{equation}
\LD{endowed with suitable initial conditions.}
After calibrating model \eqref{eq:suihtervv} on a time range $[t_0,t_0^f]$, model \eqref{eq:suihtervv_forecast} can be initialized at time $t_0^f$ using the parameters of the last phase of the calibration to forecast the epidemic evolution in the presence of an emergent variant.

Model \eqref{eq:suihtervv_forecast} is initialized at time $t_0^f$ with data from DPC if available. In order to allow the model to account for possible errors in reported data, we introduced a Gaussian error $\epsilon \sim \mathcal{N}(0,s_k^2)$ on these values, where $s_k$ is the standard deviation of the weekly average data for the compartment $k$ during the whole calibration period. The other compartments, i.e. \textit{Susceptible}, \textit{Undetected base}, \textit{Undetected variant}, \textit{Recovered}, \textit{Vaccinated with one dose}, \textit{Vaccinated with two doses}, are initialized with the values at $t_0^f$ obtained from the simulation with model \eqref{eq:suihtervv}.

The initial values for $U^b$ and $U^v$ are obtained by splitting the value of $U$ by accounting for the prevalence of the rising variant $p_v$ given by surveillance reports of  ISS \footnote{\url{https://www.epicentro.iss.it/coronavirus/sars-cov-2-monitoraggio-varianti-indagini-rapide}}.
The transmission rate for the base virus $\beta_U^b$ is the last calibrated parameter $\beta_U$ at time $t_0^f$, suitably rescaled to take into account the fact that the fraction $p_v$ of infected individuals already contracted  the virus variant. Given the increase in transmissibility $f_v$ of the spreading variant, we obtain the following estimates for the transmission rates at time $t_0^f$
\begin{align}
    \beta_U^b(\RED{t_0^f}) &= \frac{\beta_U(t_0^f)}{1+(f_v-1)p_v}, \\[3mm]
    \beta_U^v(\RED{t_0^f}) &= f_v\, \beta_U^b(t_0^f).
\end{align}

The coefficients \RED{$\sigma_1^b$ and $\sigma_2^b$ accounts for the susceptibility of vaccinated individuals w.r.t. the base version of the virus} and their value are the same as $\sigma_1$ and $\sigma_2$ introduced in Section \ref{eq:suihtervv}. The coefficients $\sigma_1^v=0.5$ and $\sigma_2^v=0.22$\footnote{\url{https://assets.publishing.service.gov.uk/government/uploads/system/uploads/attachment_data/file/1000512/Vaccine_surveillance_report_-_week_27.pdf}} instead are modified to take into account the reduced vaccines effectiveness against virus variants.
The parameters $\Tilde{\omega}_I$, $\Tilde{\omega}_H$ and $\Tilde{\gamma}_I$ are modified parameters to take into account the vaccines effect in reducing the severity of the disease and the mortality.
Based on the vaccines surveillance report of Italian\footnote{\url{https://www.epicentro.iss.it/vaccini/pdf/report-valutazione-impatto-vaccinazione-covid-19-6-ott-2021-it.pdf}} and UK governments, we defined the coefficients $h_1$, $h_2$, $t_1$, $t_2$, $m_1$, and $m_2$.
\begin{itemize}
    \item $h_1=0.655$ and $h_2=0.417$ denote the probabilities of a isolated infected individual who is vaccinated (with one or two doses, respectively) to be hospitalized w.r.t. an unvaccinated isolated infected individual;
    \item $t_1=0.63$ and $t_2=0.6$ denote the probabilities of a hospitalized individual who is vaccinated (with one or two doses, respectively) to be admitted in ICU w.r.t. an unvaccinated hospitalized individual;
    \item $m_1=0.724$ and $m_2=0.35$ denote the probabilities of an isolated infected individual who is vaccinated (with one or two doses, respectively) to die  w.r.t. an unvaccinated isolated infected individual.
\end{itemize}

\REV{Since in model~\eqref{eq:suihtervv_forecast}, the infected compartments ($U$, $I$, $H$ and $T$) collect both vaccinated and unvaccinated individuals, in order to quantify the effect of the vaccination on the worsening and mortality parameters ($\omega_I$, $\omega_H$ and $\gamma_I$), we first estimate the partition of new positives coming from either $S$, $V_1$, or $V_2$ and entering either $U_b$ or $U_v$ compartments as follows}
\begin{alignat}{3}
        u_S^b(t) &= \frac{S(t)}{N_b(t)}, \qquad u_1^b(t) &= \frac{\sigma_1^b V_1(t)}{N_b(t)}, \qquad u_2^b(t) &= \frac{\sigma_2^b V_2(t)}{N_b(t)},\label{eq:partition_no_variant}\\
u_S^v(t) &= \frac{S(t)}{N_v(t)}, \qquad u_1^v(t) &= \frac{\sigma_1^v V_1(t)}{N_v(t)}, \qquad u_2^v(t) &= \frac{\sigma_2^v V_2(t)}{N_v(t)},
\end{alignat}
where $N_b(t) = S(t) + \sigma_1 V_1(t) + \sigma_2 V_2(t)$ and $N_b(t) = S(t) + \sigma_1^v V_1(t) + \sigma_2^v V_2(t)$.

The modified worsening and mortality parameters $\Tilde{\omega}_I$, $\Tilde{\omega}_H$ and $\Tilde{\gamma}_I$ including the vaccine effects are then obtained as
\begin{align}
    \label{eq:hosp}
    \Tilde{\omega}_I(t) &= \omega_I(t_0^f)\, \frac{ u_s(t) + h_1 u_1(t) + h_2 u_2(t) }{ u_s(t_0^f) + h_1 u_1(t_0^f) + h_2 u_2(t_0^f) }, \qquad t>t_0^f,\\[1em]
    \label{eq:icu}
    \Tilde{\omega}_H(t) &= \omega_H(t_0^f) \, \frac{ u_s(t) + t_1 u_1(t) + t_2 u_2(t) }{ u_s(t_0^f) + t_1 u_1(t_0^f) + t_2 u_2(t_0^f)}, \qquad t>t_0^f,\\[1em]
    \label{eq:mort}
    \Tilde{\gamma}_I(t) &= \gamma_I(t_0^f)\, \frac{ u_s(t) + m_1 u_1(t) + m_2 u_2(t) }{ u_s(t_0^f) + m_1 u_1(t_0^f) + m_2 u_2(t_0^f) }, \qquad t>t_0^f.
\end{align}
where the total partition (including both base and variant contributions) of new positives coming from $S$, $V_1$, or $V_2$ is given by
\begin{equation}\label{eq:totalpartition}
    \begin{aligned}
        u_S(t) &= \frac{u_S^b(t) N_b(t) U^b(t) + u_S^v(t) f_v N_v(t) U^v(t)}{N_b(t) U^b(t) + N_v(t) f_v U^v(t)}\\
        u_1(t) &= \frac{u_1^b(t) N_b(t) U^b(t) + u_1^v(t) f_v N_v(t) U^v(t)}{N_b(t) U^b(t) + N_v(t) f_v U^v(t)}\\
        u_2(t) &= \frac{u_2^b(t) N_b(t) U^b(t) + u_2^v(t) f_v N_v(t) U^v(t)}{N_b(t) U^b(t) + N_v(t) f_v U^v(t)}.
    \end{aligned}
\end{equation}
\LD{We notice that, when new variants are not spreading, then} the latter equations simply reduce to Eq.~(\ref{eq:partition_no_variant}).
Moreover, we point out that the rescaling in \eqref{eq:totalpartition} using the partition at the initial time $t_0^f$ of the forecast simulation has been used to account for the fact that the values $\omega_I(t_0^f)$, $\omega_H(t_0^f)$ and $\gamma_I(t_0^f)$ have been obtained calibrating the model on data which were already affected by the vaccination campaign (until time $t_0^f$).

\RED{Although more complex models could be devised to deal with the vaccination, for instance by splitting each infected compartment ($U$, $I$, $H$, $T$) in vaccinated and unvaccinated, we preferred to keep the model as simple as possible, but still able to account for both the reduced susceptibility of vaccinated individuals, as well as the lower risk of hospitalization and death through relations \eqref{eq:hosp}-\eqref{eq:mort}.}
\RED{Moreover, in the present model we are neglecting the waning immunity over time, which has been recently highlighted in several studies \cite{Cohn2021.10.13.21264966,doi:10.1056/NEJMoa2114228,TARTOF20211407}, since the time-frame considered in the simulations (see Section \ref{sec:results}) are limited to the Summer 2021}.  

\section{Forecast Scenarios}\label{sec:scenarios}
{The COVID-19 pandemic put a heavy burden on the Italian health system and caused a huge loss of human lives. \LD{During 2020 and early 2021,} the Italian Government imposed restrictive measures to the population, in attempts to restrain and to contain the epidemic waves.} Different NPIs have been applied since the beginning of the COVID-19 epidemic, ranging \LD{from basic containment measures (such as the mandatory use of masks) to drastic ones (full lockdown).} All these interventions modified the transmissibility of the virus by reducing the number of contacts between people or the probability to contagion associated to a contact. \LD{Within epidemiological models, the effect of these NPIs can therefore be accounted for} by acting on the transmission rate $\beta_U$. We included in the model the possibility of considering different future scenarios, corresponding to different possible NPIs or social events that may also lead to a change in the transmission rate (such as, for instance, school re-openings or holidays). {We introduced a transmission coefficient $\tau$ that measures the change of the transmission rate at a given time due to imposition of restrictions (NPIs) or other social events and occurrences (schools opening, holidays, etc.). The coefficient $\tau$ is computed for the scenario to simulate, as well for the actual conditions in the last phase of the calibration; see Appendix~\ref{app:scenary}. The transmission rate $\beta_U$ is indeed} modulated by considering the change of the restriction coefficient as follows
\begin{equation}\label{eq:betatau}
    \beta_U(t) = \frac{\tau(t)}{\tau(t_0^f)}\, \beta_U(t_0^f), \qquad t>t_0^f.
\end{equation}
{We remark that the transmission coefficient $\tau$ is used in a relative fashion for predictions, i.e. changes in NPIs and social activities are encapsulated in the term $\frac{\tau(t)}{\tau(t_0^f)}$ that modulates the predicted transmission rate.}
Several scenarios can be considered, each one with its own restriction coefficient. 
\LD{Novel restrictions soon to be implemented} can be referred to population belonging to a specific age group (e.g. distance learning for schools) or to specific places (e.g. restaurants and bars closed). \LD{In order to} take into account the heterogeneity of the measures we consider the interactions between people of different age groups in different contexts and modify them according to the scenario considered.
Starting from the matrices of average number of contacts by age and context obtained from \cite{mossong2008social}, we associate a level of risk to each age group and context of exposure.
\texttt{SUIHTER} is a homogeneous model and does \LD{not explicitly accounts for age or exposition context structure,} thus the restriction coefficient is obtained from the restriction matrix by a weighted mean on the age groups with weights given by the amount of population in each group.  \REV{Examples of scenario analyses using this approach to quantify the effect of different NPIs have already been presented and discussed in \cite{parolini2021mathematical}. The procedure for computing the restriction coefficient $\tau$ for a given NPI scenario is described Appendix~\ref{app:scenary}. Additional details will be available in a forthcoming paper.}

\section{Results and discussion}\label{sec:results}
In this section we present the results obtained with the proposed model. In particular, in Section  \ref{sec:validation}, we first provide a validation of the forecast capabilities of the model by comparing with other single model forecasts as well as ensemble forecasts. The role of the new features introduced in the model, namely variants and vaccines, are investigated in the numerical results presented in Sections \ref{sec:results_variant} and \ref{sec:results_vaccine}, respectively.  Finally, in Section \ref{sec:results_scenario}, the results of a scenario analysis related to the introduction of Green Pass restrictions are presented and discussed.



\subsection{Forecast assessment}\label{sec:validation}
The SUIHTER model has been included in the European COVID-19 Forecast Hub \cite{forecasthub} which collects short-term forecasts of COVID-19 cases (i.e. new contagions) and deaths across Europe, obtained by a multitude of infectious disease modelling teams, under the coordination of the European Centre for Disease Prevention and Control.

Every week, the different teams upload their forecasts of the cumulative value of incidence cases and incidence deaths per week for one or several European countries, with a forecast horizon between 1 and 4 weeks. The forecasts are supplied in an interval format, including for each quantity point evaluation and confidence interval in 23 quantiles ranging from 0.01 to 0.99. 

For each forecast the absolute error (AE) and the weighted interval score (WIS), for both cases and deaths, are computed. The absolute error is simply the absolute value of the difference between the forecast and the measured data.
The weighted interval score is a proper scoring rule proposed in \cite{Bracher2021} to evaluate and score forecasts in an interval format. It is a generalization of the absolute error and has three components: dispersion, overprediction and underprediction. Dispersion is a weighted average of the width of prediction intervals for different levels of uncertainty, the higher the level the higher the weight. The overprediction and underprediction are penalties added whenever an observation falls outside a central prediction interval. The penalties depend on how far the observation fall from the interval and the nominal level of the interval.
For more details on these metrics, see \cite{Cramer2021.02.03.21250974}.

The number of forecasts supplied by the different teams may be different since some of them joined the project later than others, or  forecasts in certain weeks for some models may be missing.
In order to allow a fair comparison, in \cite{Cramer2021.02.03.21250974} a relative measure of forecast performance, called \textit{relative WIS} (relWIS), was introduced, taking into account that different teams may not cover the same forecast targets. The relative WIS is obtained performing a \textit{pairwise comparison tournament} in which a score ratio between each pair of models is computed on the set of shared targets. The relative WIS is then obtained as the geometric mean of these ratios and normalized with the corresponding quantity computed for a reference baseline model. The forecast of the baseline model is defined with median given by the data measured in the most recent week, with uncertainty around the median based on changes in the past of the time series (see \cite{Cramer2021.02.03.21250974}). 

Models with smaller values of relWIS are thus performing better and a value below one means that the model performs better than the baseline model. The relWIS indicator can then be used to score different models.

The same procedure can be used to define a second scoring rule, the \textit{relative absolute error} (relAE), in which the pairwise comparison tournament is performed based on the absolute error instead of the weighted interval score (see \cite{Cramer2021.02.03.21250974}).

The 1-week horizon forecast obtained with the SUIHTER model has been uploaded weekly on the European COVID-19 Forecast Hub \cite{forecasthub} since April 2021. The performance metrics over the entire forecast history (32 weeks from May 2, 2021 to December 11, 2021) are reported in Table \ref{tab:hub_all}. 

\begin{table}[b]
\centering
\begin{tabular}{@{}ccccc@{}}
\toprule
 & \multicolumn{2}{c}{Cases} & \multicolumn{2}{c}{Deaths} \\
Model                 & relAE & relWIS & relAE & relWIS  \\ \midrule
EuroCOVIDhub-baseline & 1   & 1 &  1 & 1 \\
EuroCOVIDhub-ensemble & 0.59  & 0.55 & 0.48 & 0.42   \\
SUIHTER               & 0.31  & 0.47 & 0.29 & 0.36  \\
 \bottomrule
\end{tabular}
\caption{Relative absolute error and relative weighted interval score for the SUIHTER, baseline and ensemble forecast (with 1-week horizon) over the entire forecast history (32 weeks from May 2, 2021 to December 11, 2021)}
    \label{tab:hub_all}
\end{table}

The table reports a comparison between the 1-week forecast obtained with the SUIHTER model, the baseline forecast and the \texttt{EuroCOVIDhub-ensemble} forecast obtained using all the forecast upload on the hub for each target. By definition, the baseline forecast has a relWIS and relAE equal to one. As discussed in the introduction, ensemble forecasts are usually expected to be more accurate and robust than those produced by single models.
The results presented in Table \ref{tab:hub_all} 
clearly show that the SUIHTER model is able to produce short-term forecasts which outperforms the ensemble forecasts for both cases and deaths.   Moreover, the results reported for Italy on \cite{forecasthub} show that, according to the relWIS scoring indicator, the SUIHTER model ranked second on cases forecast (behind the MUNI-ARIMA model and second on deaths forecast (behind the LANL-GrowthRate	model) among 24 models. 


For the 10 forecasts in the time range from August 28, 2021 to November 2, 2021, multiple horizon forecasts (ranging from 1 to 4 weeks) have also been evaluated and compared with the \texttt{EuroCOVIDhub-ensemble}.  The results are presented in Table \ref{hub_4weeks_last10}. As expected both the relative weighted interval score and the relative absolute error increase as longer time horizon are considered.  It is worth noticing that the good performances of the SUIHTER model are confirmed for short term predictions of cases forecast (up 2-weeks horizons), where the SUIHTER results are found to be better than both the baseline and ensemble forecasts in terms of absolute error and weighted interval score.  Moreover, superior performances of the SUIHTER model for deaths forecast are found even for longer time horizons (up to 4 weeks), where very low values of both relAE and relWIS are obtained.

\begin{table}[]
\centering
\begin{tabular}{@{}cccccc@{}}
\toprule
 & & \multicolumn{2}{c}{Cases} & \multicolumn{2}{c}{Deaths} \\
Model & Horizon & relAE & relWIS & relAE & relWIS \\ \midrule
\multirow{4}{*}{EuroCOVIDhub-ensemble} 
& 1 & 0.72 & 0.69 & 0.78 & 0.40   \\
& 2 & 1.02 & 0.93 & 0.59 & 0.32  \\
& 3 & 1.16 & 1.00 & 0.52 & 0.30  \\
& 4 & 1.35 & 1.15 & 0.64 & 0.38  \\
 \midrule
\multirow{4}{*}{SUIHTER} 
& 1  & 0.62 & 0.60 & 0.73 & 0.36 \\
& 2  & 0.93 & 0.89 & 0.49 & 0.27\\
& 3  & 1.13 & 1.21 & 0.48 & 0.28\\
& 4  & 1.21 & 1.39 & 0.49 & 0.31\\
 \bottomrule
\end{tabular}
\caption{Relative absolute error and relative weighted interval score for the SUIHTER and ensemble forecast  (with 1 to 4-week horizons) over the last 10 weeks (from August 28, 2021 to October 30, 2021)}
    \label{hub_4weeks_last10}
\end{table}

\begin{figure}
\centering
\subcaptionbox{2021-09-06}[.32\linewidth]{\includegraphics[width=\linewidth,trim={20 20 20 20},clip]{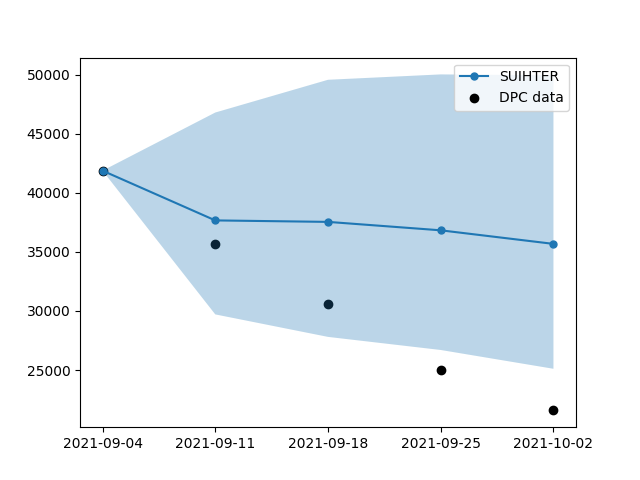}}
\subcaptionbox{2021-09-13}[.32\linewidth]{\includegraphics[width=\linewidth,trim={20 20 20 20},clip]{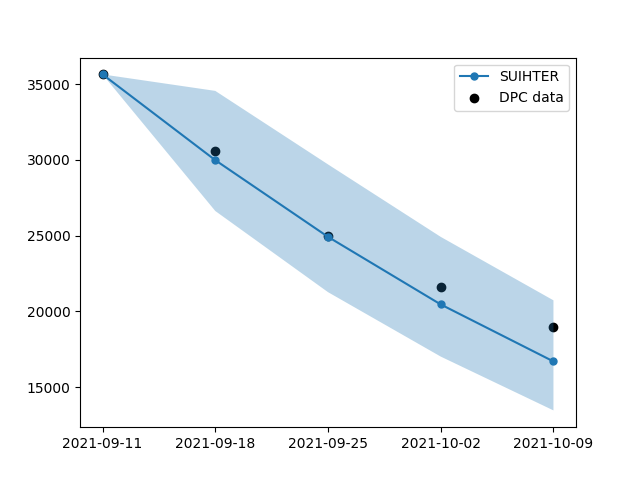}}
\subcaptionbox{2021-09-20}[.32\linewidth]{\includegraphics[width=\linewidth,trim={20 20 20 20},clip]{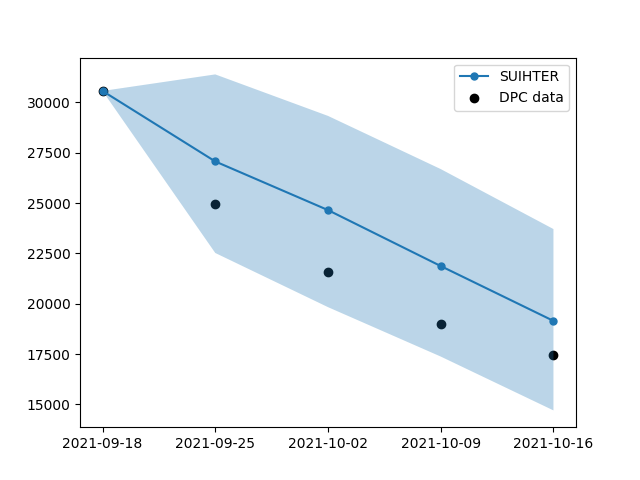}}\\
\subcaptionbox{2021-09-27}[.32\linewidth]{\includegraphics[width=\linewidth,trim={20 20 20 20},clip]{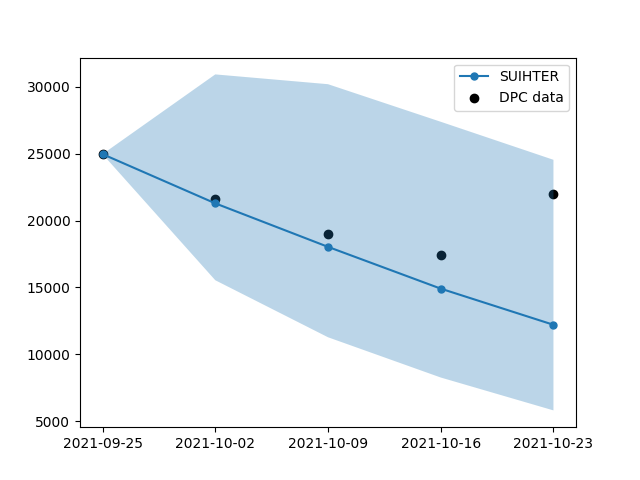}}
\subcaptionbox{2021-10-04}[.32\linewidth]{\includegraphics[width=\linewidth,trim={20 20 20 20},clip]{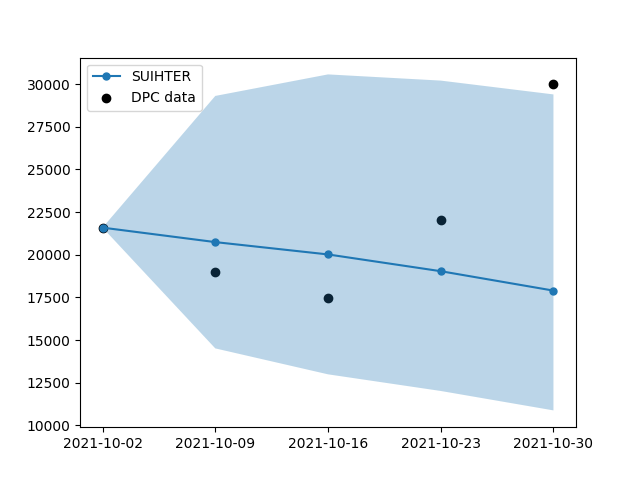}}
\subcaptionbox{2021-10-11}[.32\linewidth]{\includegraphics[width=\linewidth,trim={20 20 20 20},clip]{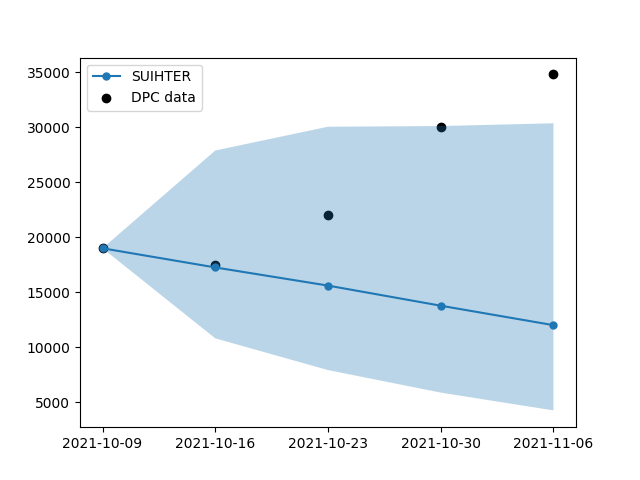}}\\
\subcaptionbox{2021-10-18}[.32\linewidth]{\includegraphics[width=\linewidth,trim={20 20 20 20},clip]{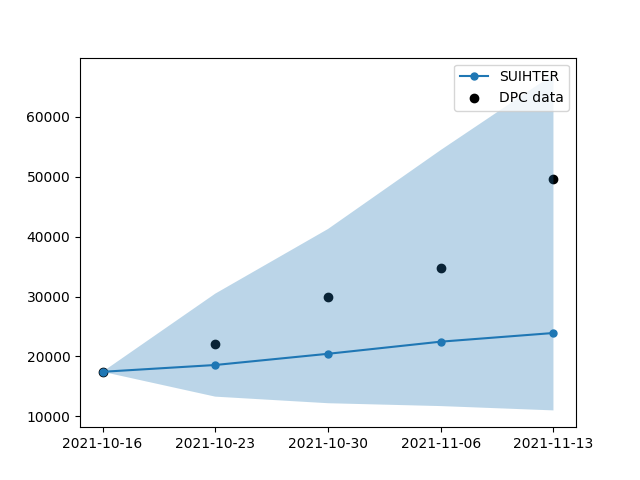}}
\subcaptionbox{2021-10-25}[.32\linewidth]{\includegraphics[width=\linewidth,trim={20 20 20 20},clip]{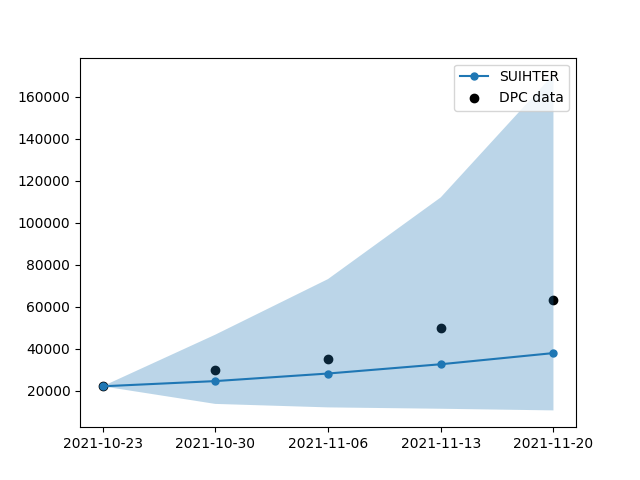}}
\subcaptionbox{2021-11-01}[.32\linewidth]{\includegraphics[width=\linewidth,trim={20 20 20 20},clip]{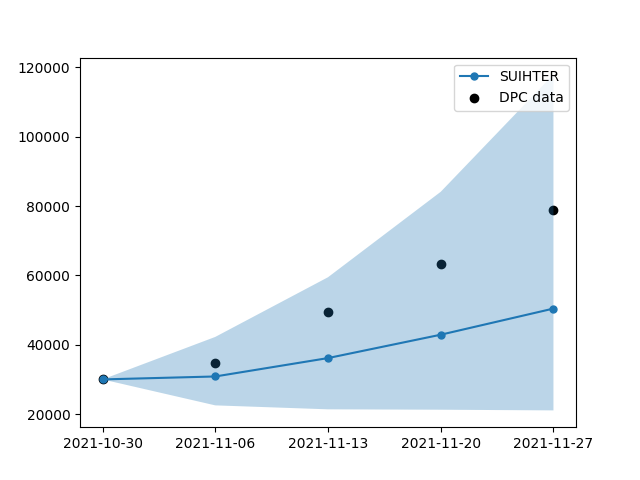}}
\caption{Weekly new cases forecasts computed with the SUIHTER model on a 4-weeks horizon with 95\% confidence interval along with actual data provided by DPC.}
\label{fig:forecast_cases}
\end{figure}

\begin{figure}
\centering
\subcaptionbox{2021-09-06}[.32\linewidth]{\includegraphics[width=\linewidth,trim={20 20 20 20},clip]{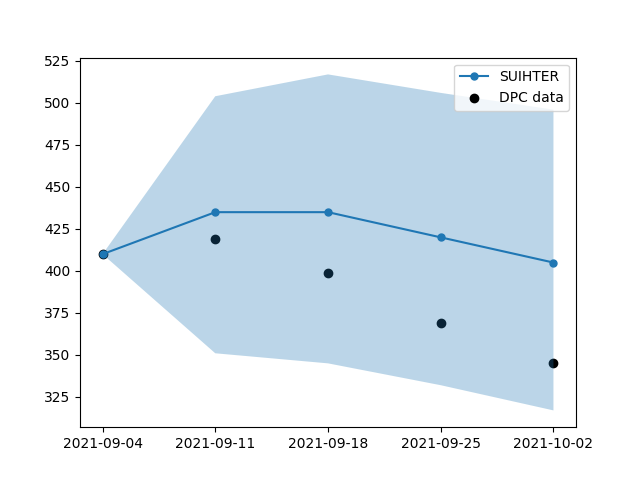}}
\subcaptionbox{2021-09-13}[.32\linewidth]{\includegraphics[width=\linewidth,trim={20 20 20 20},clip]{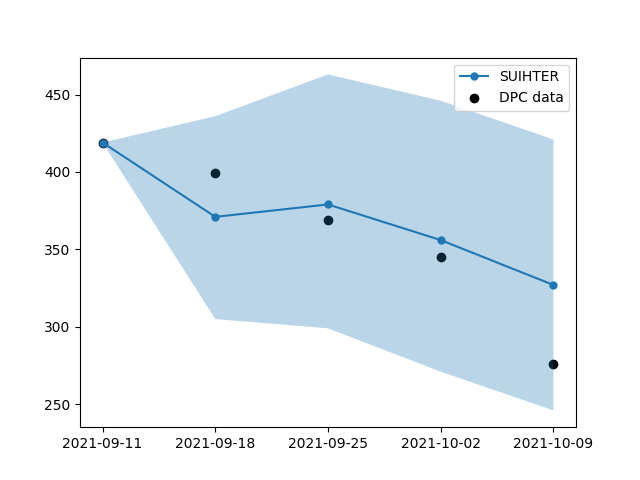}}
\subcaptionbox{2021-09-20}[.32\linewidth]{\includegraphics[width=\linewidth,trim={20 20 20 20},clip]{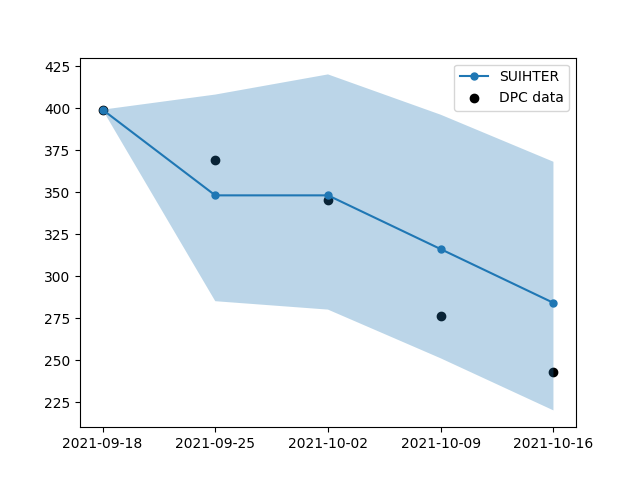}}\\
\subcaptionbox{2021-09-27}[.32\linewidth]{\includegraphics[width=\linewidth,trim={20 20 20 20},clip]{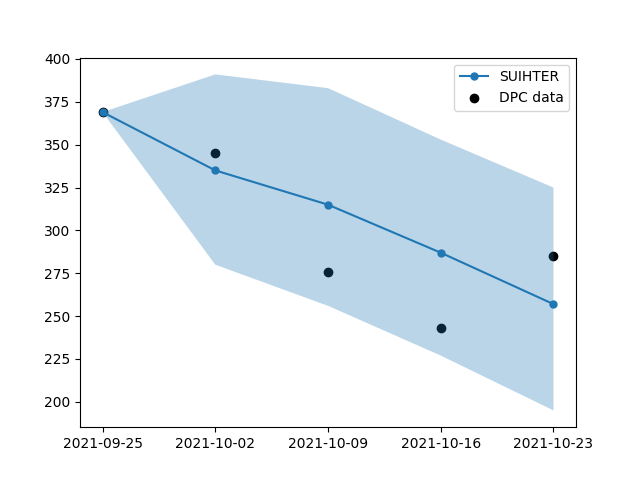}}
\subcaptionbox{2021-10-04}[.32\linewidth]{\includegraphics[width=\linewidth,trim={20 20 20 20},clip]{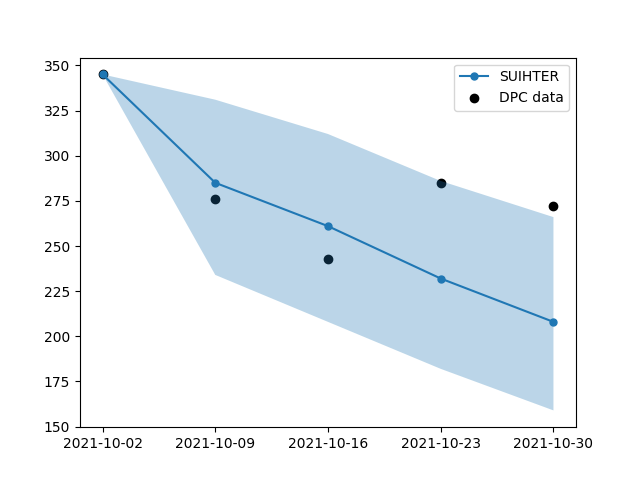}}
\subcaptionbox{2021-10-11}[.32\linewidth]{\includegraphics[width=\linewidth,trim={20 20 20 20},clip]{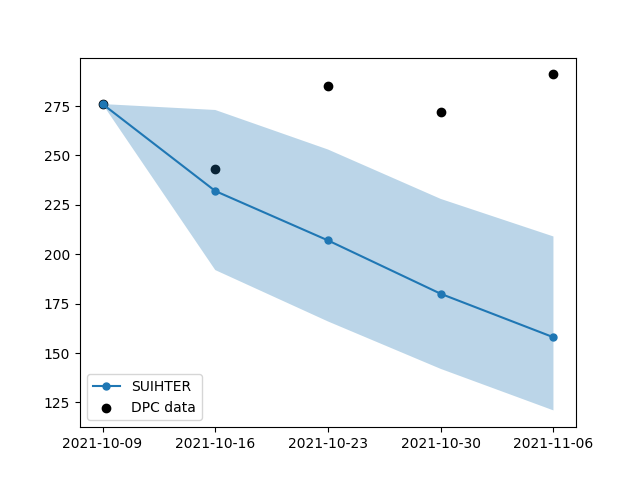}}\\
\subcaptionbox{2021-10-18}[.32\linewidth]{\includegraphics[width=\linewidth,trim={20 20 20 20},clip]{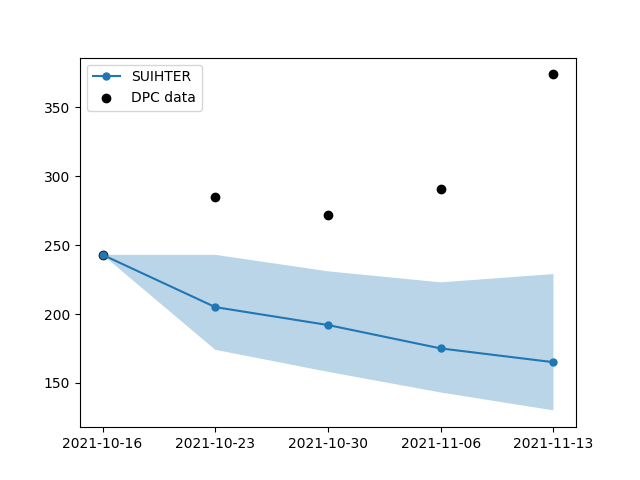}}
\subcaptionbox{2021-10-25}[.32\linewidth]{\includegraphics[width=\linewidth,trim={20 20 20 20},clip]{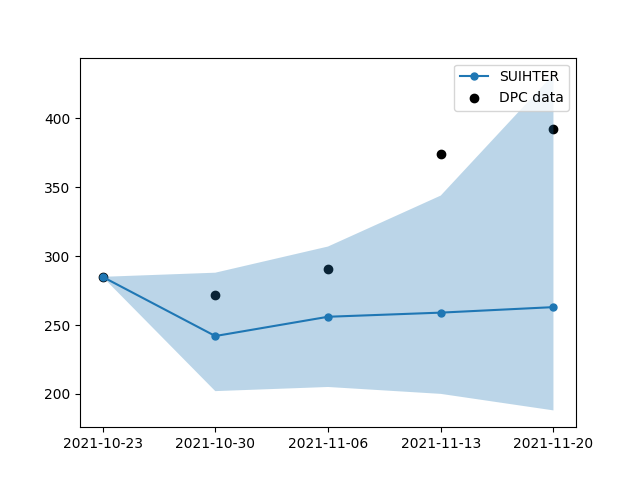}}
\subcaptionbox{2021-11-01}[.32\linewidth]{\includegraphics[width=\linewidth,trim={20 20 20 20},clip]{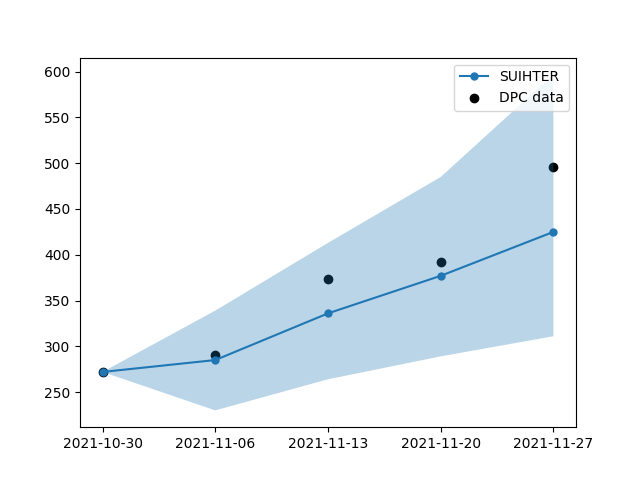}}
\caption{Deaths forecasts computed with the SUIHTER model on a 4-weeks horizon with 95\% confidence interval along with actual data provided by DPC.}
\label{fig:forecast_deaths}
\end{figure}

In Figures \ref{fig:forecast_cases} and \ref{fig:forecast_deaths} we report the forecasts with the predicted median value and 95\% prediction interval along with the actual data of cases and deaths, respectively. The dates on the x-axis denote the target end date, i.e. the last day of the week for which the forecast has been made. \REV{We can note that for the cases forecast the data fall within the prediction intervals (with only 2 exceptions for longer-time forecasts). Concerning deaths forecasts, the model fails to supply a reliable forecast in one case (see Figure \ref{fig:forecast_deaths} (f)). However, we should point out that the reported data for the week ending on October 23, 2021 were biased by the large number of deaths (28) not previously reported that have been added that week to correct the data.}

\subsection{Effect of new variants}\label{sec:results_variant}
\begin{figure}[hb!]
\centering
\includegraphics[width=\textwidth, trim={120 70 120 70}, clip]{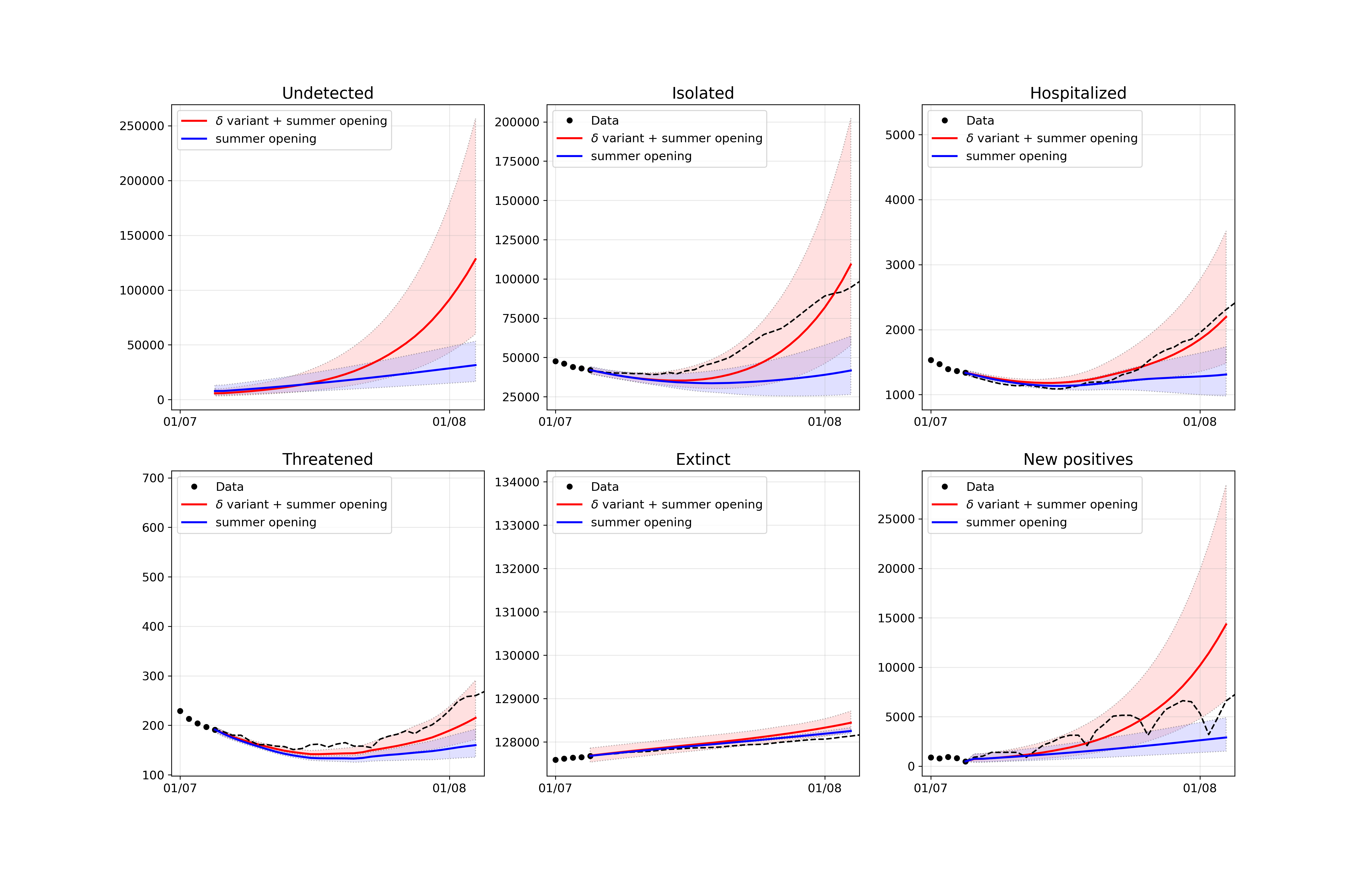}
\caption{Simulations performed from 5 July to 5 August with (red) and without (blue) taking into account the spreading of the highly infectious \textit{Delta} variant}
\label{fig:delta_variant}
\end{figure}

During the spreading of a new variant it is fundamental to consider that it is way more transmissible than the prevalent one.  For this reason we tried to predict the evolution of the epidemic when the \textit{Delta} variant was spreading in Italy in July 2021, eventually becoming the prevalent one.
We performed two simulations, the first one with model (\ref{eq:suihtervv_forecast}) described in section \ref{sec:variant_model}, initialized with prevalence data provided by surveillance reports of ISS \footnote{\url{https://www.epicentro.iss.it/coronavirus/sars-cov-2-monitoraggio-varianti-indagini-rapide}}. The second simulation was made with $p_v=0$, thus assuming that no new variant was spreading. In this case the calibrated transmission rate $\beta_U$ is not rescaled but used as is. Both  simulations start from July 5, 2021 and last for two months until August 5, 2021. The parameters have been calibrated taking into account data available from February 20, 2021 to July, 5 2021. 
The two simulations are shown in Figure \ref{fig:delta_variant} where it is possible to appreciate that considering the \textit{Delta} variant the model was able to predict the rise of the epidemic curve occurred in July, which would not be predicted if the role of the variant would have been neglected. 

\begin{figure}[hb!]
\centering
\includegraphics[width=\textwidth, trim={120 70 120 70}, clip]{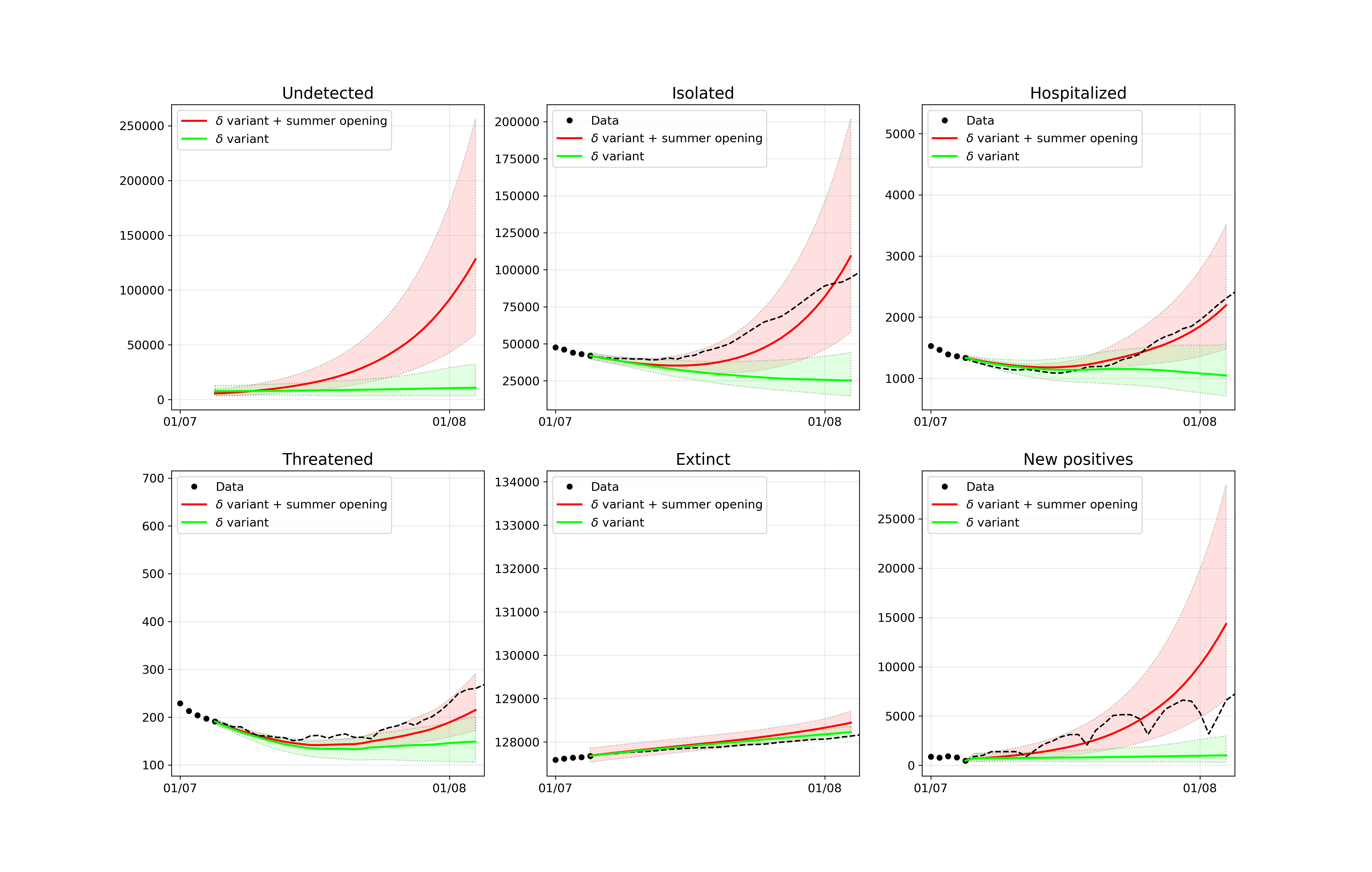}
\caption{Simulations performed from 5 July to 5 August with (red) and without (green) taking into account the summer openings resulting in an increased number of social contacts}
\label{fig:summer_open}
\end{figure}

Note that this case features the effect of summer holidays with people gathering in tourism regions and contracting the virus. {This was accounted for in the simulations by introducing a summer scenario as detailed in Section ~\ref{sec:scenarios}, resulting in a $42.7 \%$ increase on $\beta_U$, i.e. such that $\frac{\tau(t)}{\tau(t_0^f)} = 1.427$ in Eq.~\eqref{eq:betatau}.} 
Should we have neglected the spreading of the \textit{Delta} variant (see Figure~\ref{fig:delta_variant}), the summer re-opening scenario alone would not be enough to justify the number of cases observed.
Similarly, we simulated for the same period, with the same set of parameters, but without accounting for the summer scenario, thus considering only the effect of the \textit{Delta} variant. The result is shown in Figure \ref{fig:summer_open} alongside with the simulation including both the \textit{Delta} variant and the summer scenario.
Also in this case, the \textit{Delta} variant alone could not make the epidemic curve rise as it has been observed, thus implying that the phenomenon witnessed in July was very likely due to a combination of the spreading of the \textit{Delta} variant as well as the growth in the number social contacts due to summer vacations.

\subsection{Effects of the vaccination}\label{sec:results_vaccine}
The evolution of the COVID-19 epidemic in 2021 has been strongly influenced by the introduction of effective vaccines. In the present section we present a set of numerical results to illustrate how the extended SUIHTER model introduced in Section \ref{sec:model} can improve the forecast quality. To show the effect of the presence of vaccinated compartments, we simulated the trend of the epidemic starting from May 22, 2021 and lasting for two months until July 22, 2021.
The parameters are obtained by calibrating the SUIHTER model on the DPC data from February 20, 2021 to May 22, 2021 including the vaccinated compartments too. Then, the model has been recalibrated over the same period of time, on the same data, but neglecting the presence of vaccinated compartments (setting to zero both $v_1$ and $v_2$ as well as the initial values of $V_1$ and $V_2$). Even though the two simulations yield almost the same values for most of the compartments, the one without vaccines produces a higher number of deaths. This is due to the fact that it includes the observed effect of vaccines directly in the calibrated transmission rates so obtaining a similar result on positive cases but it does not consider the mortality reduction given by vaccines, which is instead considered in model (\ref{eq:suihtervv_forecast}). Furthermore, model (\ref{eq:suihtervv_forecast}) can give us an estimation of how positives split into unvaccinated, partially vaccinated with one dose or fully vaccinated with second dose.
 
\begin{figure}
\centering
\includegraphics[width=\textwidth, trim={120 70 120 70}, clip]{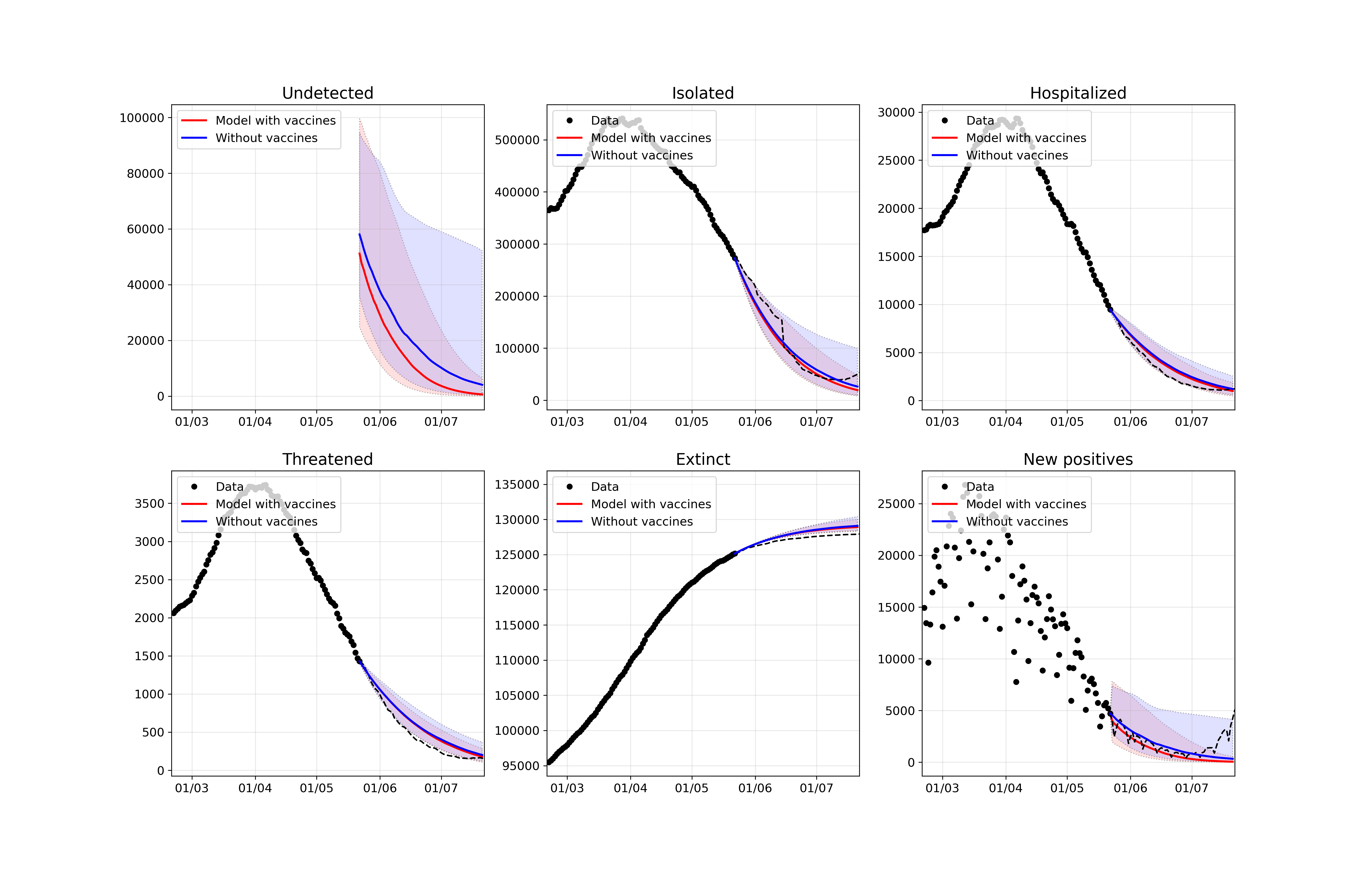}
\caption{Comparison between the models with vaccines (red) and without vaccines (blue} in a forecast analysis covering the time interval between May 22, 2021 and July 22, 2021.
\label{fig:no_vaccines}
\end{figure}
\begin{figure}
\centering
\includegraphics[width=\textwidth, trim={110 0 110 0}, clip]{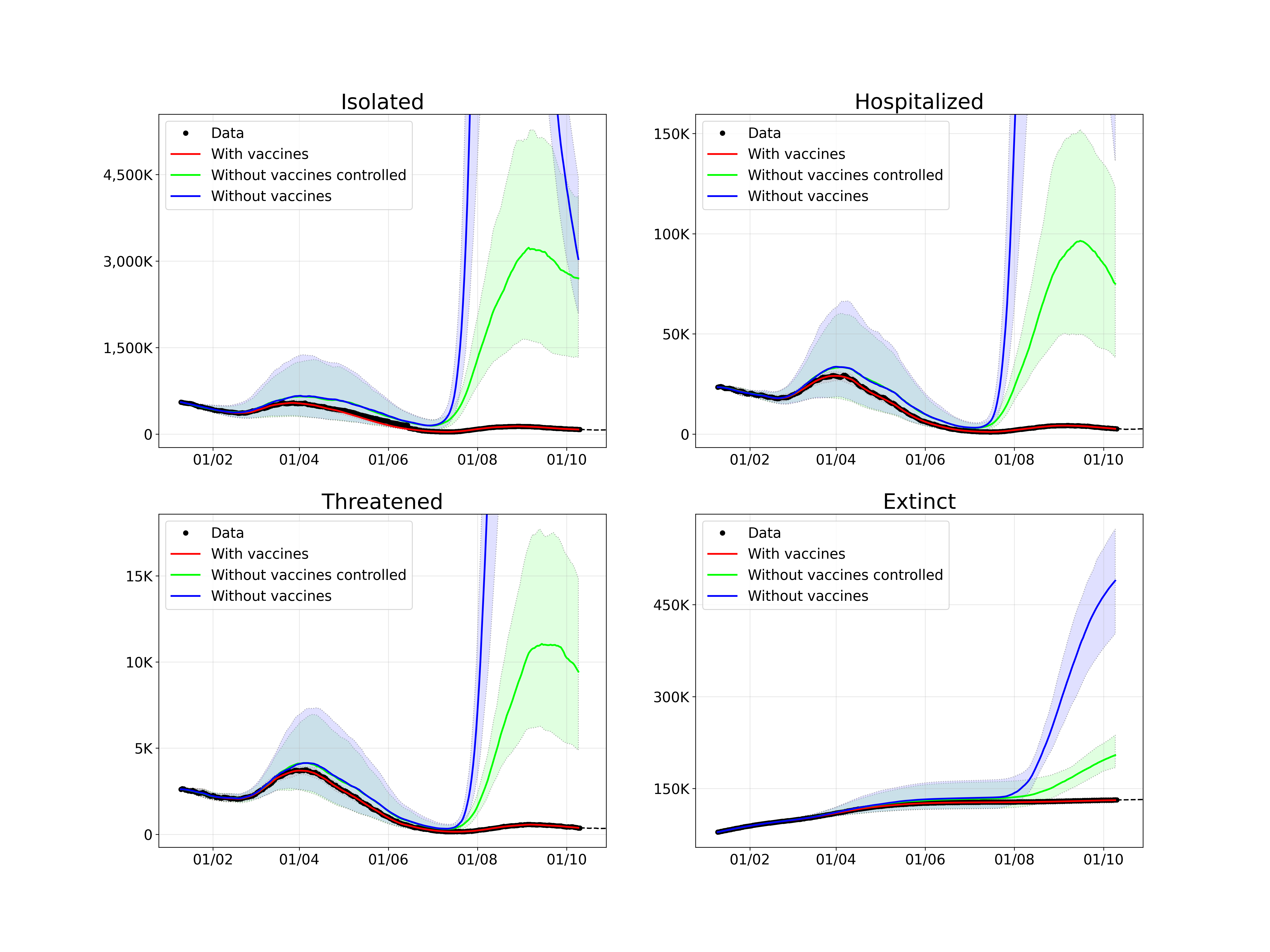}
\caption{Simulations of the epidemic during vaccination campaign with what-if scenarios neglecting the presence of vaccinated compartments}
\label{fig:vaccini_secondo}
\end{figure}

To show the effect of the vaccines on the model and estimate how the epidemic would have evolved without the vaccines, we compared three simulations obtained with the same set of parameters calibrated from January 10, 2021 to  October 10, 2021. The first simulation is obtained with model~(\ref{eq:suihtervv}) described in Section~\ref{sec:model}, used also for calibration, and thus taking into account vaccination compartments and reproduces faithfully the epidemic history. The simulation is then repeated with the same set of parameters removing the vaccinated compartments. In Figure~\ref{fig:vaccini_secondo}, we see that the epidemic curves have an uncontrolled exponential increase from the beginning of July, corresponding to summer openings and spreading of the \textit{Delta} variant. Obviously this scenario in unrealistic as some containment measures would have been applied before reaching a critical level of new cases and hospital beds occupancy. The third analyzed scenario (in green in Figure~\ref{fig:vaccini_secondo}) considers the imposition of automatic restrictions when critical incidence levels are reached. Three thresholds are set, based on the guidelines prescribed by the Italian government\footnote{Decreto Legge 23 luglio 2021, n. 105, https://www.gazzettaufficiale.it/eli/id/2021/09/30/21A05687/sg}. The first restrictions are the ones imposed in yellow zones and are applied once the weekly incidence is above 50 cases per 100\,000 population. When the incidence reaches 150 cases  per 100\,000 population, orange zone restrictions are applied and finally red zone restrictions are imposed once the limit of 250 cases per 100\,000 population has been exceeded. 
Even though in this scenario the number of cases has been greatly reduced, the containment measures are not enough to prevent a big epidemic wave in the summer. 
Analyzing these scenarios we can speculate on how things would have gone without the vaccination campaign and on how many human lives have been saved.

\subsection{Scenario analysis: adoption of Green Pass}\label{sec:results_scenario}
\LD{Since August 6, 2021, people are required to hold the so-called Green Pass for certain activities, a document that certifies to have been vaccinated, having been tested negative in the previous 48 hours, or recovered from COVID-19. The Green Pass started to be required for accessing indoor restaurants, long-distance travel, and  cultural and leisure activities.} Starting September 1, 2021, the Green Pass has been also required to all the personnel working at school (from grade 1 to universities) as well as to university students. More recently, on October 15, 2021, this rule has been extended to all working categories. The objective of the measure is clearly to control the spread of the epidemic and to encourage a further extension of the vaccination campaign to those people which are not yet vaccinated.

In this context, the SUIHTER model has been applied to estimate the effect of the introduction of the Green Pass at school, with the two-fold objectives:
\begin{enumerate}
    \item estimating whether the school re-opening in mid September 2021 would have produced a new epidemic wave (as happened in September 2020);
    \item quantifying at which extent a strict control on the Green Pass certification would have been critical to guarantee a control of the epidemic.
\end{enumerate}

\begin{figure}
\centering
\includegraphics[width=\textwidth, trim={120 70 120 70}, clip]{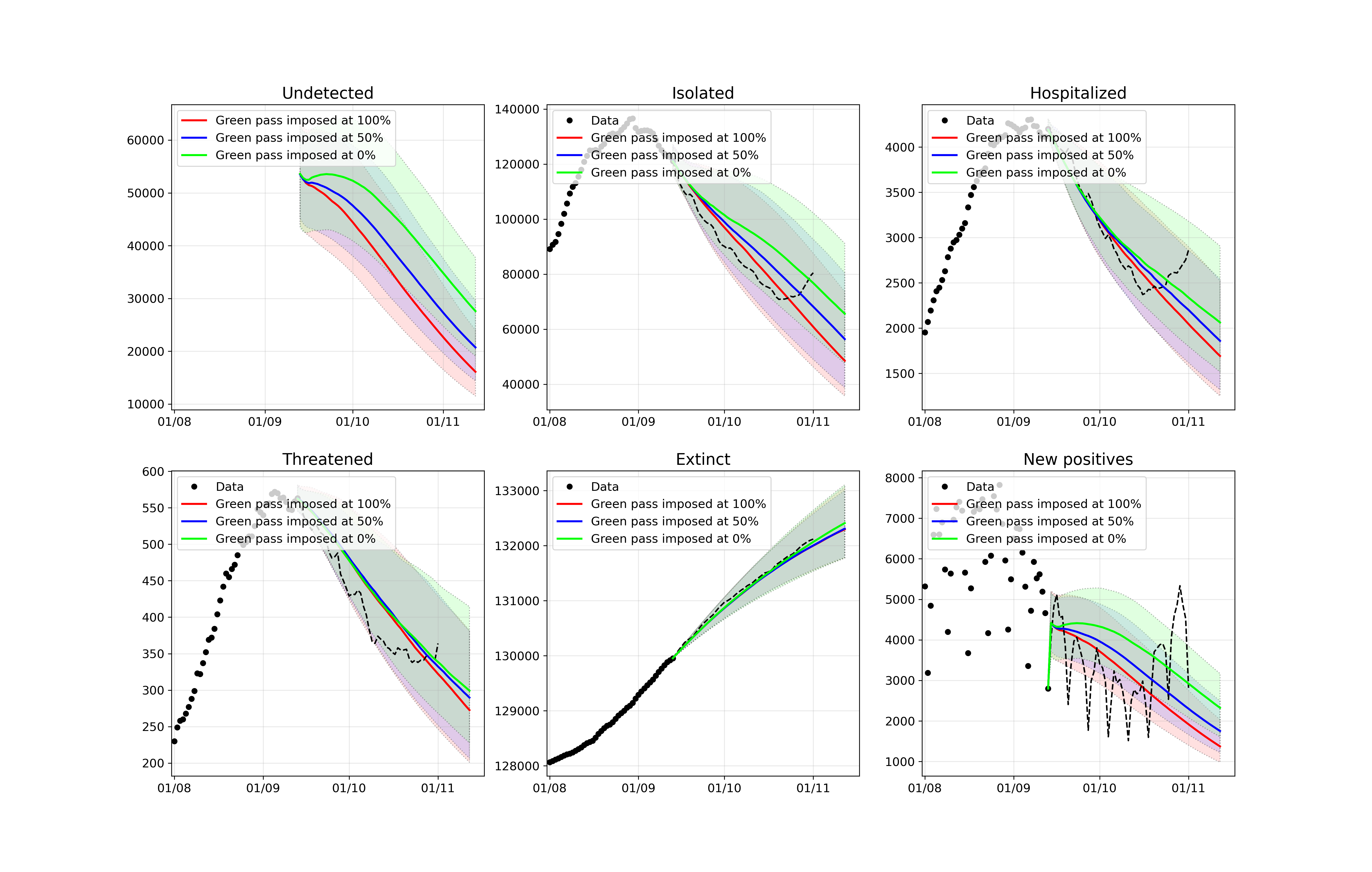}
\caption{Scenario analysis comparing the effect of different levels of imposition of Green Pass in schools}
\label{fig:green_pass}
\end{figure}

A scenario analysis has been carried out calibrating the SUIHTER model with the data available until September 10, 2021 and performing a two-months forecast until November 10. In order to account for the effect of the adoption of the Green Pass in schools, we computed two different contact matrices for each scenario: one for the vaccinated people (assuming all the people having the Green Pass from vaccination only) and the other for susceptible individuals. From these matrices, we obtained two different restriction coefficients, leading to two \LD{$\beta_U$ rates,} one for the compartment $S$, accounting for unvaccinated people, and thus without Green Pass and one for $V_1$ and $V_2$.
We considered three different scenarios with increased $\beta_U$ due to school re-openings, assuming the Green Pass is made compulsory for high schools and university students and staff:
\begin{enumerate}
    \item the Green Pass is strictly enforced so that unvaccinated individuals are not entering schools and universities;
    \item \REV{the Green Pass is mildly enforced resulting in 50\% of unvaccinated individuals still entering schools and universities};
    \item there are no controls on Green Pass so that everyone can enter schools and universities, even though they are not vaccinated.
\end{enumerate}
The results are shown in Figure~\ref{fig:green_pass}. In none of the three cases analysed the increase of $\beta_U$ is enough to produce a new epidemic wave. However, there is a significant difference on the number of positive cases. \RED{This difference may have been important in view of the new epidemic wave that occurred in Italy (as in the rest of Europe during the Fall 2021, since it allowed to start from a lower incidence level the new exponential growth phase.}

\section{Conclusions} \label{sec:conclusions}
In this paper, we have presented a new epidemiological model derived from the original SUIHTER model introduced in \cite{parolini2021suihter} for the analysis of the COVID-19 epidemic in Italy. The range of applicability of the model has been extended to cover new aspects of the epidemic that became indispensable to be considered for an accurate description of the phenomenon, namely the emergence of new virus variants and the role of the vaccination campaign. 

In particular, the new model accounts for the presence of different virus variants to describe the transition phases in which the wild-type virus (or the variant which is predominant at a given time) is replaced by a new variant characterized by a higher transmission rate. This feature has to be accounted for in order to capture variant outbreaks, such as the one observed in Italy in Summer 2021 when the Delta variant increased its prevalence from 22\% to 95\% in one month.

The variant model can also be combined with the vaccine model by considering different transmission rate reduction factors (which define to the efficacy of the vaccination in reducing the transmission) for wild type virus or specific variants. 

The reduced transmission is not the only way vaccines contributes in the model. The worsening rates also account for the increasing number of vaccinated individuals, by including the consequent reduced probability of developing a serious disease once vaccinated.  

The role of the vaccination in controlling the evolution of the epidemic in Italy during the Spring and Summer 2021 has been quantified through a \textit{what-if} scenario analysis. The results indicate that, without the vaccination campaign, the Summer outbreak (mostly due to the Delta variant) would have stroke Italy with an unprecedented force and, even with strict NPIs restrictions, a new epidemic wave, even stronger than the previous ones, would have likely occurred.

The reliability of the results produced by the SUIHTER model, in particular concerning forecast analysis, have also been investigated.
The competitive performances of the SUIHTER model in short-term forecasting have been quantitatively verified by comparison with other single model forecasts that joined the European COVID-19 Forecast Hub project \cite{forecasthub}. These results highlight the capability of our model to supply reliable near-future forecasts of cases up to two weeks horizon), while longer-term horizon scenarios (up to four weeks) can be successfully covered when considering forecasts of deaths, thanks to the rich compartmental structure of the SUIHTER model which includes the number of individuals who are hospitalized and hosted in ICUs.

\REV{The limitation of the original SUIHTER model, which are shared by several models involving a large number of compartments, were discussed in details in \cite{parolini2021suihter}. Most of them, in particular the homogeneous nature of the model which does not include spatial heterogeneity neither age stratification, also apply to the extension presented in the present paper. The range of applicability of the new model is broader since it allows to simulate and forecast the epidemic evolution in the presence of virus variants and vaccines. The proposed approach to deal with emerging virus variants is rather general and may applied to any new variant of concern for which an estimate on the increased transmission rate (relative to previous variants) is available.  \LD{The vaccination model can} be further extended to include the reduction of the vaccine efficacy over time  and the additional \textit{booster} dose that should be able to guarantee a longer vaccine immunity \cite{doi:10.1056/NEJMoa2114255}}.


\section*{Acknowledgements}
The authors would like to thank Prof. Andrea Pugliese for his insightful suggestions and careful reading of the manuscript. The authors acknowledge Giulia Villani for the help provided in the setup of the numerical simulations. This research has been partially funded by \textit{Dipartimento per le politiche della famiglia, Presidenza del Consiglio dei Ministri}, under the Agreement "Un modello matematico per lo studio dell’epidemia da COVID19 su scala nazionale” (DIPOFAM 0000192 P-4.26.1.9 del 15/01/2021).


\begin{table}[t!]
\centering
\begin{tabular}{|c|c|c|c|c|c|c|c|c|}
\hline
\textbf{Ages}   & \textbf{Total} & \textbf{Family} & \textbf{Home} & \textbf{School} & \textbf{Work} & \textbf{Transport} & \textbf{Leisure} & \textbf{Other}  \\
\hline
\textbf{0-4}    & 16.54 & 2.30  & 2.19  & 5.27   & 0.00    & 0.98      & 3.06    & 2.75 \\
\hline
\textbf{5-9}    & 20.49 & 2.27  & 2.34  & 8.87   & 0.00    & 1.12      & 4.53    & 1.37 \\
\hline
\textbf{10-14}  & 27.38 & 2.21  & 2.22 & 11.98  & 0.05  & 1.35      & 5.62    & 3.80  \\
\hline
\textbf{15-19}  & 29.28 & 2.05  & 2.54 & 13.22  & 0.05 & 1.74      & 6.83    & 2.87 \\
\hline
\textbf{20-24}  & 22.15 & 1.49  & 2.02  & 1.17   & 4.49 & 0.96      & 7.23    & 4.80  \\
\hline
\textbf{25-29}  & 21.00 & 1.04  & 2.43  & 2.23   & 5.21 & 1.13      & 6.30    & 2.66 \\
\hline
\textbf{30-34}  & 18.03 & 1.26  & 2.29  & 0.85   & 3.92 & 0.76      & 5.24    & 3.72 \\
\hline
\textbf{35-39}  & 21.25 & 1.75  & 2.63  & 0.68   & 7.78 & 1.05      & 3.92    & 3.45 \\
\hline
\textbf{40-44}  & 22.35 & 1.63  & 2.25  & 2.53   & 7.00    & 0.67      & 4.48    & 3.79 \\
\hline
\textbf{45-49}  & 19.27 & 1.50  & 1.49  & 2.61   & 8.24 & 0.88      & 1.93    & 2.64 \\
\hline
\textbf{50-54}  & 22.30 & 1.38  & 1.37  & 5.54   & 8.05 & 0.52      & 2.02    & 3.41 \\
\hline
\textbf{55-59}  & 18.27 & 1.11  & 1.77  & 1.41   & 4.60  & 0.68      & 3.62    & 5.06 \\
\hline
\textbf{60-64}  & 18.43 & 0.91  & 2.37  & 1.07   & 6.05 & 0.87      & 3.53    & 3.63 \\
\hline
\textbf{65-69}  & 12.74 & 0.71  & 2.39  & 0.55   & 0.48 & 0.95      & 3.33    & 4.33 \\
\hline
\textbf{70+}    & 10.55 & 0.71  & 2.53  & 0.06   & 1.04 & 0.22      & 4.22    & 1.77 \\
\hline
\end{tabular}
\caption{Average number of contacts in Italy by age group and social context of exposition.}
\label{table:context}
\end{table}

\appendix
\section{Determining the transmission coefficient for NPI scenarios}\label{app:scenary}
The transmission coefficient $\tau$ introduced in Section~\ref{sec:scenarios} (Eq.~\eqref{eq:betatau}) encapsulates in a single scalar value the effect of different NPIs (distancing measures, restriction on leisure activities, distance learning) or social occurrences and events (school opening and closing, school breaks and holidays) that may be accounted in forecasting the evolution of the epidemics. Since the NPIs are often involving specific age-groups -- like schools activities -- as well as specific contexts of exposition, we evaluate the transmission coefficient $\tau$ based on a multi-age multi-group characterization of the contact matrix. 
We consider the POLYMOD contact matrices for Italy proposed in \cite{mossong2008social}, counting the average number of contacts that individuals from different age-groups (0-4, 5-9, 10-14, 15-19, 20-24, 25-29, 30-34, 35-39, 40-44, 45-49, 50-54, 55-59, 60-64, 65-69, 70+ for a total of $N_a=15$ groups) have in the different contexts: home, school, work, transport, leisure and other. We remark that, with respect to the standard table, we use data on the composition of households obtained from ISTAT (for 2019) \cite{istat}, to further split the contacts taking place at home into a \textit{family} context (in which contacts between family members are counted) and a \textit{house} context (in which contacts occurring at home among individuals who are not family members are counted); see \cite{villani}. This 
has the advantage to account for NIPs limiting visits of non-family members to other households, a restriction that was often enforced in Italy during 2020 and early 2021. 
As example, we limit here to report in Table~\ref{table:context} the average number of contacts by age group in the $N_c=7$ contexts of exposition.

Let us indicate by $c^k_{ij}$ the (average) number of contacts of individuals in age group $i$ with individuals in age group $j$ in the context of exposition $k$, where $i,j=1,\ldots,N_a=15$ and $k=1,\ldots,N_c=7$. Then, Table~\ref{table:context} represents $\widetilde c^k_i = \sum_{j=1}^{N_a} c^k_{ij}$, i.e. the average number of contacts of individuals in age group $i$ with other individuals in the context of exposition $k$.
The transmission coefficient from individuals in the age group $i$ to other individuals in context of exposition $k$ reads
$$
\widetilde \tau^k_{i} = \widetilde c^k_{i} \, p_i^k,
$$
for $i=1,\ldots,N_a=15$ and $k=1,\ldots,N_c=7$, where $p_i^k$ is the probability that a contact of a infectious individual in age group $i$ is able to expose any other individual in context $k$. Specifically, we determine $p_i^k$ as
$$
p_i^k = r_{a,i} \, f^k_{i}\left( r_{c,i}^k, s^k_i \right),
$$
where: $r_{a,i}$ is the transmission risk associated to age group $i$ ($r_{a,i}=0.34$ for $i=1,2,3$, $r_{a,i}=1$ for $i=4,\ldots,13$, and $r_{a,i}=1.47$ for $i=1,2,3$ \cite{CTS}); $r_{c,i}^k$ is the risk of transmission associated to individuals in the age group $i$ within the exposition context $k$ (for example contacts occurring in context $k=1$ ``family" are riskier than those occurring in $k=4$ ``work"); $s^k_i\in [0,1]$ accounts for NPIs and social occurrences for individuals in age group $i$ and in context $k$ (for example, during school holidays $s^3_i=0$ for all $i=1,\ldots,N_a=15$, instead if NPIs prescribe students in high schools attending only at $50\%$ in presence, then $s^3_i=0.5$); the functions $f^k_i$ combine the coefficients $r_{c,i}^k$ and $s^k_i$ based on age group $i$ and context $k$. These functions are designed to account for different subcontexts of exposition within a given $k$, like it occurs for $k=6$ (``leisure") in restaurants, bars, sports, etc... or for $k=4$ (``work") for different working environments as healthcare, manufacturing, etc... The function $f^k_i$ is also used to account for the mismatch between age groups and grades in school or other contexts; for example, NIPs regarding schools are prescribed in Italy based on grades, not age. For more details on the definition of the transmission coefficient $\widetilde \tau_{i}^k$ we refer the interested reader to \cite{villani}.

Finally, the transmission coefficient $\tau$ used in Eq.~\eqref{eq:betatau} is obtained as
$$
\tau = \sum_{k=1}^{N_c} \sum_{i=1}^{N_a} \varsigma_i \, \widetilde \tau_{i}^k, 
$$
where $\varsigma_i$ is the percentage of Italian population within the age group $i$ \cite{istat}.

\bibliography{SUIHTER.bib}

\end{document}